\documentclass[12pt,a4paper]{article}
\usepackage[a4paper,top=28mm]{geometry}

\usepackage[utf8]{inputenc}
\usepackage[T1]{fontenc}
\usepackage{tabularx,xcolor,graphicx}
\usepackage{multirow}
\usepackage{booktabs}
\usepackage{pdfpages}

\usepackage[space]{grffile}

\newcommand{\opname}[1]{\mathop{\mathrm{#1}}\nolimits}
\newcommand{\Pf}{\opname{Pf}}   

\usepackage{nicefrac}
\usepackage{textcomp}
\usepackage{tcolorbox}
\usepackage{cancel}
\usepackage{bbold}

\providecommand{\si}[1]{#1}

\providecommand{\measuredangle}{\mathord{\angle}}

\providecommand{\eV}{\mathrm{eV}}

\newcommand{\SOC}{\mathrm{SOC}}
\newcommand{\ZM}{\mathrm{ZM}}
\newcommand{\Imag}{\mathrm{Im}}
\providecommand{\text}[1]{\textrm{#1}}

\usepackage{float}
\usepackage[toc,page]{appendix}

\usepackage[backend=biber, sorting = none, citestyle = numeric-comp, maxbibnames=10]{biblatex}
\usepackage{blindtext}
\usepackage{xcolor}
\usepackage{listings, lstautogobble}
\usepackage{subcaption}

\newcommand{\LL}{\mathbf{L}}
\newcommand{\VS}{\mathbf{S}}

\newcommand{\kk}{\mathbf{k}}

\usepackage[breaklinks]{hyperref}

\title{Quantum Spin Hall Phase in the Truncated Trihexagonal Lattice:\\
  A Topological Archimedean Structure}

\author{L.~V.~Duc Pham$^{1,2}$, Nicki F.~Hinsche$^{1}$, Ingrid Mertig$^{1}$}
\date{}

\begin{document}

\maketitle

\begin{center}
$^{1}$Institute of Physics, Martin Luther University Halle-Wittenberg, 
06120 Halle (Saale), Germany\\
$^{2}$Faculty of Chemistry and Food Chemistry, Technical University Dresden, 
01062 Dresden, Germany\\[1ex]
Corresponding author: \texttt{phleviet@mpi-halle.mpg.de}
\end{center}

\begin{abstract}
Archimedean lattices constitute a unique family of two-dimensional tilings formed from regular polygons arranged with uniform vertex configurations. 
While the kagome and snub square lattices, the simplest members of the Archimedean lattice family, have been extensively investigated — the former as a paradigmatic system for geometric frustration and nontrivial band topology, and the latter primarily as a quasicrystal approximant — the broader family remains largely unexplored in terms of electronic and topological properties.
In this work, we present a systematic Python-based tight-binding study of all eight pure Archimedean lattices, modeled as two-dimensional carbon-based networks serving as a proof-of-principle system.
We analyze their band structures, investigate topological edge states arising from unconventional nanoribbon geometries, and evaluate $\mathbb{Z}_2$ invariants as well as intrinsic spin Hall conductivities using the Kubo formalism.
Our results reveal that several Archimedean lattices, such as the truncated hexagonal and truncated trihexagonal lattices, host nearly dispersionless flat bands extending across the Brillouin zone, which remain robust even in the presence of next-nearest-neighbor hopping and strong spin–orbit coupling. In particular, the truncated trihexagonal lattice supports topologically protected, highly spin-polarized edge states across multiple ribbon geometries. These states are stable against defects and spin-flip scattering, and give rise to quantized spin Hall currents.
\end{abstract}

\pagebreak
\section{Introduction}
Two-dimensional (2D) materials with nontrivial band topology have emerged as a fertile ground for realizing exotic electronic phases, such as the quantum spin Hall (QSH) effect and topological insulators. A key focus in this domain is the manipulation of spin currents without magnetic fields, which is crucial for low-dissipation spintronics applications. The intrinsic spin Hall effect, driven by spin-orbit coupling in systems with time-reversal symmetry, enables the generation of spin-polarized edge currents protected from backscattering \cite{KaneMelePRL2005,BernevigScience2006}.

Among 2D systems with nontrivial band topology, kagome lattices have drawn particular attention due to their unique geometric frustration and flat band characteristics, which can host correlated and topologically nontrivial states and might realize topological magnetism and superconductors with exotic properties \cite{YeNature2018,YinNature2019,Kato2024,PhysRevMaterials.5.034801,PhysRevB.111.214412}.
Further prominent 2D lattices, e.g. the snub square lattice and elongated triangular lattice, have been recently studied as approximants for quasicrystalline structures \cite{SteinhardtOstlund1987,Bechinger_quasiC,RoyMohseni2016,VoigtQuasiC20,Shi20242464} and as artificial geometries in photonic and electronic metamaterials \cite{Jovanovic_photonics,Stelson_photonics,Chelnokov_photonics}. 
All these lattices belong to a broader family of Archimedean tilings, distinguished by their vertex-transitive symmetry and variety of polygonal motifs. Despite the extensive focus on the kagome lattice, other members of the Archimedean family, such as the snub square, elongated triangular, truncated hexagonal or truncated trihexagonal lattice, remain underexplored regarding their potential for hosting topological non-trivial electronic phases.

The history of Archimedean lattices dates back to 1619, when Kepler first systematically described and identified all possible two-dimensional tilings of regular polygons with identical local environments \cite{kepler1997harmony}. In total, eleven such uniform tilings exist: three regular lattices—composed of a single polygon type, including the triangular, square, and hexagonal lattices (the latter realized in the prominent material graphene)—and eight semiregular, or mathematically Archimedean, lattices that combine multiple polygon types in a vertex-transitive arrangement. These eight pure Archimedean lattices, which are the focus of this paper, have since become a cornerstone in geometry, crystallography, and, more recently, condensed matter physics. The Archimedean lattices diverse coordination environments lead to rich edge terminations and the possibility of engineering distinct topological edge states. Recent studies suggest that the interplay of lattice geometry, orbital hybridization, and spin-orbit coupling in these systems could facilitate novel quantum Hall phases beyond those seen in conventional honeycomb or square lattices.

Nowadays several members of the Archimedean lattice family could, in principle, be realized experimentally using recent advances in surface assembly and molecular design. Artificial electronic lattices formed from CO molecules on (111) metal surfaces have been demonstrated by scanning tunnelling microscopy (STM) manipulation \cite{Gomes2012,GhaemiCOsurface}. 
Epitaxial deposition of Bi atoms on Tl/Si(111) surfaces has successfully yielded quasi-periodic tiling structures exhibiting Archimedean-lattice-like features \cite{GruznevTlBi}.
Related bottom-up approaches include metal–organic frameworks (MOFs) \cite{ChenMOF21,VoigtQuasiC20,Shi20242464} and halogen hydrogen-bonded organic frameworks (XHOFs)\cite{YinXOFS}, which enable the creation of a wide variety of two-dimensional network topologies. Flat polycyclic molecules such as triangulene derivatives, as well as on-surface polymerization methods, further extend the possibilities for realizing covalently bonded two-dimensional lattices \cite{hongdeTriangulene,hongdeTriangulene2,Niu2Dpoly}.

2D carbon-based materials seem to provide an ideal platform to explore the physics of Archimedean lattices \cite{Lu_2D_allotropes,paupitz_review_allotr}. The most elementary member, graphene, realizes the regular hexagonal lattice, while kekulene \cite{Zhang_kekulene} represents a finite molecular analogue of the same topology. In recent years, several two-dimensional carbon allotropes -- such as biphenylene \cite{FanBiphenylene}, graphyne \cite{DesyatkinGraphyne}, and graphdiyne \cite{Li_graphdiyne,Serafini_graphdiynes} -- have been successfully synthesized, all composed of networks of carbon polygons that closely resemble certain Archimedean lattice topologies. Graphenylene \cite{Song_Graphenylene}, consisting of four-, six-, and twelve-membered rings, has been identified as a close realization of the truncated trihexagonal lattice. 
The structural simplicity of these systems -- built solely from carbon atoms -- and the tunable hybridization between $sp$ and $sp^2$ orbitals make two-dimensional carbon networks an appealing platform for studying the electronic and topological properties of Archimedean lattices.

In this work, we employ a Python-based tight-binding framework of two-dimensional carbon networks to investigate the electronic structure of all eight pure Archimedean lattices. We analyze their topological characteristics through the $\mathbb{Z}_2$ invariant and evaluate their intrinsic spin Hall conductivities using the Kubo formalism, in order to assess their potential for spintronic device applications. Furthermore, we examine the topological edge states that emerge in nanoribbons with unconventional edge terminations, highlighting how different boundary geometries affect their dispersion and spatial localization while preserving the underlying bulk topology. The source code and illustrative examples are available as open-source resources \cite{zenodo_pham_hinsche_Mertig_2025}.

\section{Methodology}
\subsection{Setting up the lattices}
\label{sec:method_ase}

\begin{figure}[ht]
    \centering
    \includegraphics[width=0.4\linewidth]{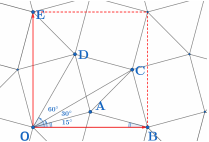}
    \caption{The unit cell of the snub square lattice}
    \label{fig:snub_square_unit_cell}
\end{figure}

Setting up the lattices is the first essential step in performing calculations. In our code, this is done within the framework of the \textsc{python} package \textsc{ASE} \cite{ase-paper}. First, we need to identify the position of each site in the unit cell as well as the lattice vectors for every lattice. This task, although being a task of elementary Euclidean geometry, can be quite complicated. Thus, we provide a brief geometrical demonstration on the snub square lattice (Fig.~\ref{fig:snub_square_unit_cell}). In this case, one possible way to choose the unit cell would be the square defined by the two vectors \textbf{OB} and \textbf{OE}. The basis of the unit cell will then be the points O, A, C, and D. Since all triangles are
equilateral, their interior angles are $60^{\circ}$, and therefore the angle $\measuredangle OAB$ is $150^{\circ}$. The triangle $\triangle{OAB}$
is isosceles because all edges in Archimedean tilings are equal. Hence, the angle $\measuredangle AOB$ is
$\frac{1}{2}(180^{\circ} - 150^{\circ}) = 15^{\circ}$. Point A is therefore obtained by rotating point $A_1$ with coordinates $A_1(a, 0, 0)$
by $15^{\circ}$. Analogously, we conclude that $\measuredangle OAC =  150^{\circ}$, which implies $\measuredangle AOC =  15^{\circ}$.
Consequently, $OB$ is equal to $OC$. $OB$ can be calculated from the law of cosines as $a\sqrt{2 + \sqrt{3}}$.
The point $C$ is obtained by rotating point B with coordinates $B(a\sqrt{2 + \sqrt{3}}, 0 , 0)$ by an angle of $30^{\circ}$. From this calculation, we also conclude that the lattice constant is $a\sqrt{2 + \sqrt{3}}$.

We can determine $\measuredangle DOB$ as the sum of $\measuredangle AOB$ and $\measuredangle DOA$. Thus, it is
$60^{\circ}$ - note that $\measuredangle DOA$ is the angle between the edge of a square and its diagonal and is therefore $45^{\circ}$.
The segment $OD$ is the diagonal of a square with a side length of $a$ and is consequently $a\sqrt{2}$. The point $D$ can then be obtained by
rotating the point $D_1$ with coordinates $(a\sqrt{2}, 0, 0)$ by an angle of $60^{\circ}$. The sites positions and the lattice vectors can be used as input to create an instance of the object \texttt{ase.Atoms()} from \textsc{ase}. Afterwards, this instance, together with a matrix deciding the size of the structure, can be passed to the method \texttt{build.make\_supercell()} to create the desired planar structure. All Archimedean lattices presented in this paper are made available by the authors as Crystallographic Information Files (CIFs) via an open data repository \cite{zenodo_pham_hinsche_Mertig_2025}. 
For the 2D bulk calculation, the input for \texttt{build.make\_supercell()} is given by the structure defined via \texttt{ase.Atoms()} and the unity matrix. 
We note that the term bulk, here and throughout the manuscript, refers to the periodic, edge-free two-dimensional system, i.e., the ideal infinite lattice used for bulk band-structure calculations. The information on the structures such as the orbitals' positions, lattice vectors and neighboring sites of a given site within a certain radius can then be extracted as output.
By adjusting the periodic boundary condition with the \texttt{pbc} argument of the object \texttt{ase.Atoms()}, i.e. keeping the periodicity in only one direction, and changing the size of the structure by adjusting the matrix input for \texttt{build.make\_supercell()}, 1D nanoribbons are generated in a straightforward manner.

\subsection{Tight-Binding formalism}
In our study, we consider a planar network of carbon atoms arranged according to the respective Archimedean lattice geometries. This choice represents the simplest and most transparent proof-of-principle platform. The use of carbon atoms ensures planarity and realistic bond connectivity, consistent with the known stability of many carbon-based two-dimensional allotropes \cite{Lu_2D_allotropes,paupitz_review_allotr}. Within this framework, we perform electronic-structure calculations using a tight-binding model that includes both s and p orbitals, employing the standard Slater–Koster parameters of graphene \cite{PhysRevB.82.245412}. This approach allows us to isolate and analyze the intrinsic electronic and topological properties of the Archimedean lattices without introducing material-specific complications. 
Our model includes both nearest neighbor (NN) and next nearest neighbor (NNN) hopping.
The NNN hopping amplitude is scaled by $\exp\left(-3\,\frac{d_{NNN}}{d_{NN}}\right)$,
where $d_{NN}$ and $d_{NNN}$ are the distances to nearest and next nearest neighbors, respectively.
The model also allows atomic spin-orbit coupling (SOC) as
$\mathcal{H}_{\SOC} = \lambda_{\SOC}\,\LL \cdot \VS$,
where $\lambda_{\SOC}$ is the SOC strength and $\LL$ and $\VS$ are the orbital angular momentum
and spin operators, respectively \cite{JAFFE1987399}.

In all of our calculations, to emphasize the effect of SOC, $\lambda_{\SOC}$ is always set to $1 \ \si{\eV}$. Using the structures' information generated by the method described in section \ref{sec:method_ase} and enabling electronic structure calculations via the Python package \textsc{pythtb} \cite{pythtb}, we set up and solve the tight-binding Hamiltonian for all eight Archimedean lattices as well as their various nanoribbon structures. The source code and illustrative examples are available as open-source resources \cite{zenodo_pham_hinsche_Mertig_2025}.

A complementary and elegant approach was very recently presented by Joseph and Boettcher \cite{Joseph2025}, who constructed Archimedean lattices from a statistical-physics and graph-theoretical perspective, deriving analytic results for their band structures and densities of states. While their study does not address edge properties or topological transport aspects, it offers an insightful alternative route to exploring the electronic characteristics of these lattices.

\section{Electronic structure}
\subsection{Bulk band structures}

\begin{figure}[ht]
    \centering
    \includegraphics[width=0.5\linewidth]{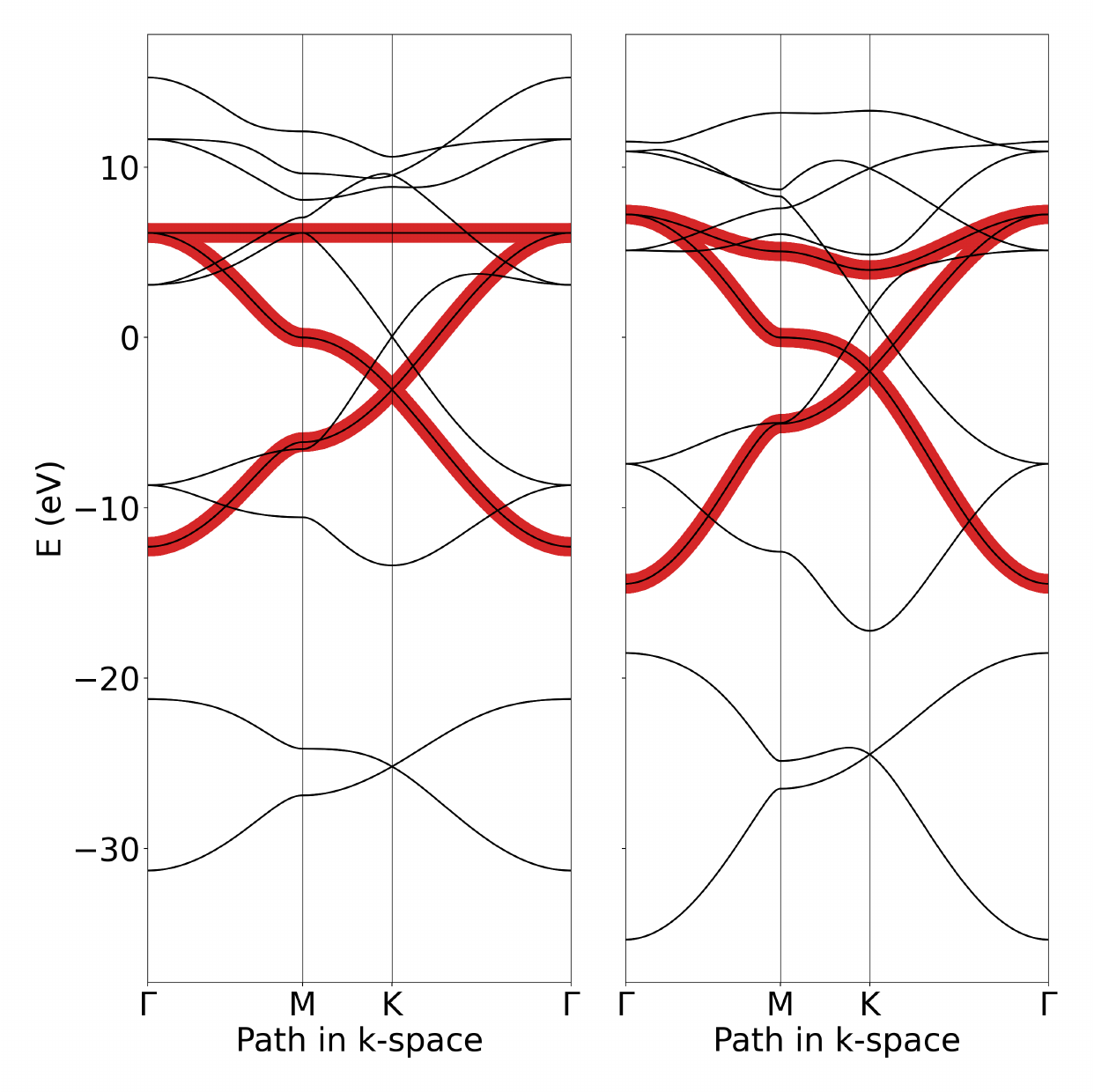}
         \caption{Band structure of the kagome lattice, including NN hopping terms (left) and both NN and NNN (right) hopping terms, the $p_z$ projected bands are highlighted in red}
    \label{fig:band_structure_kagome_projected_pz}
\end{figure}

One key feature of graphene is that its $p_z$ orbitals are decoupled from the other
orbitals within the two-dimensional plane, allowing electrons to move freely, leading to graphene's high charge carrier mobility. Therefore, it is often sufficient to focus on the $p_z$ orbitals near the Fermi level. For this reason, many works on two-dimensional materials only discuss the $p_z$ orbitals in a simple tight-binding model with one orbital per site (two orbitals when spinfull) and hopping amplitudes of $t = \pm 1$ \cite{de2019topological, springer2020topological, PhysRevB.80.113102, PhysRevLett.95.146802, PhysRevB.82.085310}. As a proof of concept, we calculate the band structure of the well-studied kagome lattice with $s$ and $p$ orbitals and performed $p_z$ orbitals projection (Fig.~\ref{fig:band_structure_kagome_projected_pz}). As expected, in  calculations with NN hopping terms only, the projected bands of $p_z$ orbitals are a perfect mirror image of the well-known band structure of the kagome lattice \cite{PhysRevB.80.113102, FAN202330, OliverBuschOH}. Consequently, the interesting electronic properties of the kagome lattice, i.e. the fully flat band and the Dirac point near the Fermi level, are maintained. Beside the Dirac point in band block with $p_z$ characteristics, there exists another Dirac point in the low energy band block. Notably, when NNN hopping terms are included, the fully flat band of the kagome lattice gains dispersion.

It is well established that spin--orbit coupling (SOC) is the key interaction giving rise to the quantum spin Hall phase in solids. Since the intrinsic SOC of carbon is rather weak, inducing and controlling SOC in carbon-based materials is essential for realizing topological states of matter and spintronic functionalities. Among the various approaches, placing carbon networks in proximity to transition-metal dichalcogenides (TMDCs) has proven to be the most effective strategy to significantly enhance SOC via interfacial hybridization~\cite{AvsarSOC,PhysRevB.111.205415,SunSOC,FabianSOC}. 
Following this concept of proximity-induced SOC, we assume a coupling strength of $\lambda_{\SOC} = 1\,\si{\eV}$ throughout our calculations to clearly illustrate the associated band inversions and topological features in a proof-of-principle manner. For comparison, we include in the Supplementary Information~\cite{SI} additional band structures calculated for smaller, yet finite, SOC values. The qualitative features---in particular, the trends on band inversion and nontrivial topology---remain present, although the inverted gaps become correspondingly smaller and thus more difficult to resolve visually.

\begin{figure}[ht]
    \begin{subfigure}{0.32\linewidth}
        \centering
        \includegraphics[width=\linewidth]{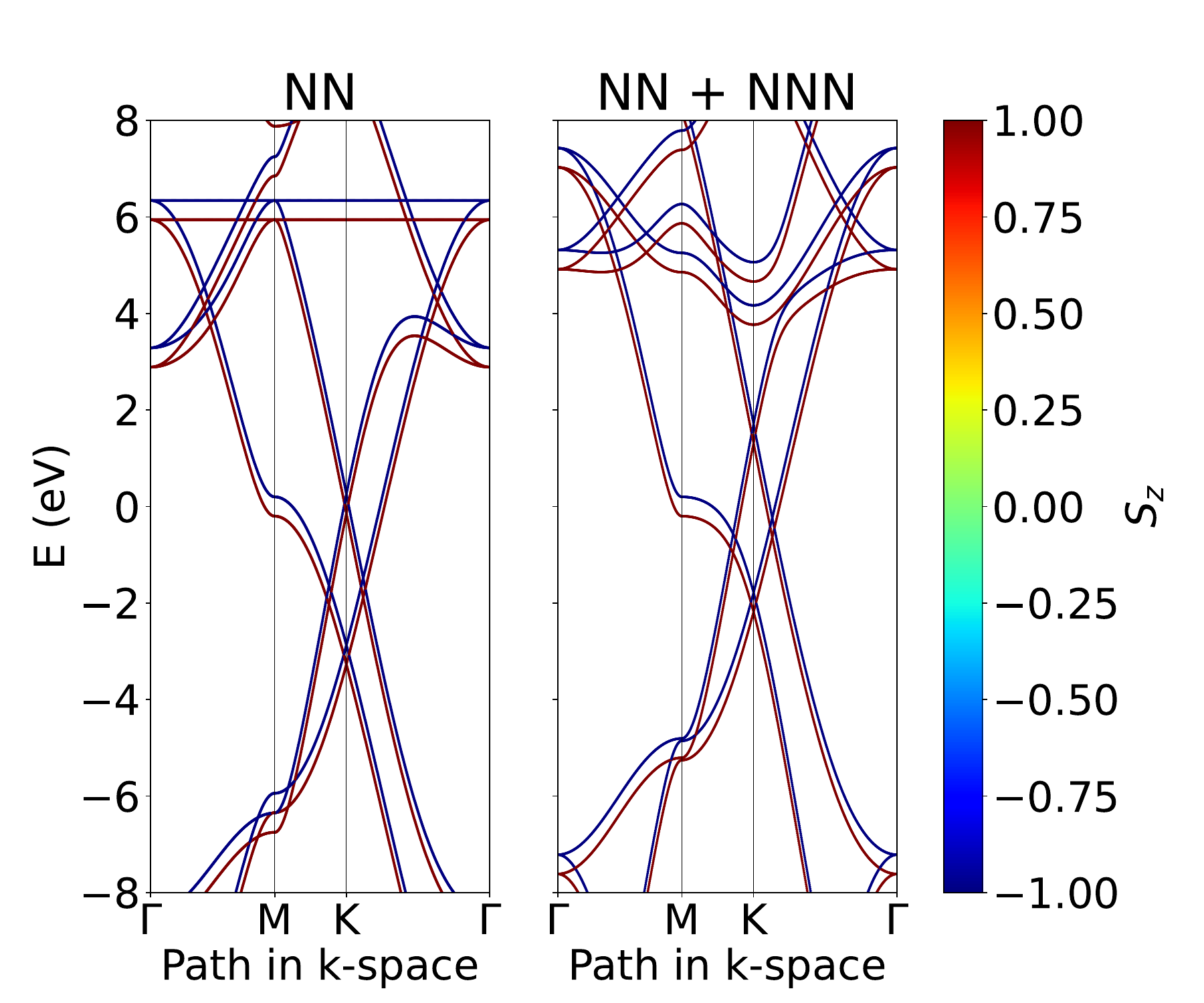} 
		\caption{}
        \label{fig:only_Zeeman_kagome}
    \end{subfigure}
    \begin{subfigure}{0.32\linewidth}
        \centering
        \includegraphics[width=\linewidth]{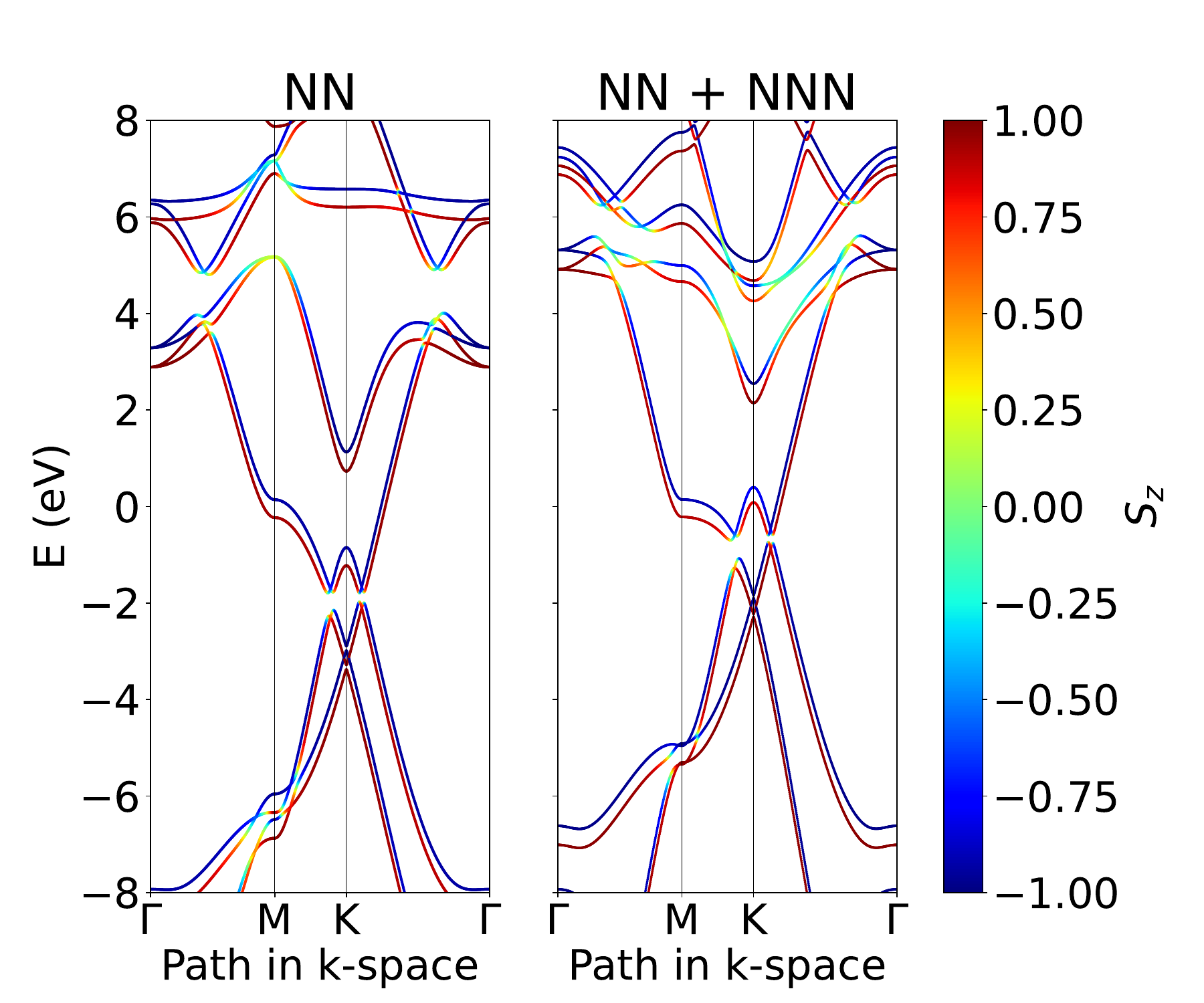} 
		\caption{}
        \label{fig:Zeeman_SOC_kagome}
    \end{subfigure}
    \begin{subfigure}{0.32\linewidth}
        \centering
        \includegraphics[width=\linewidth]{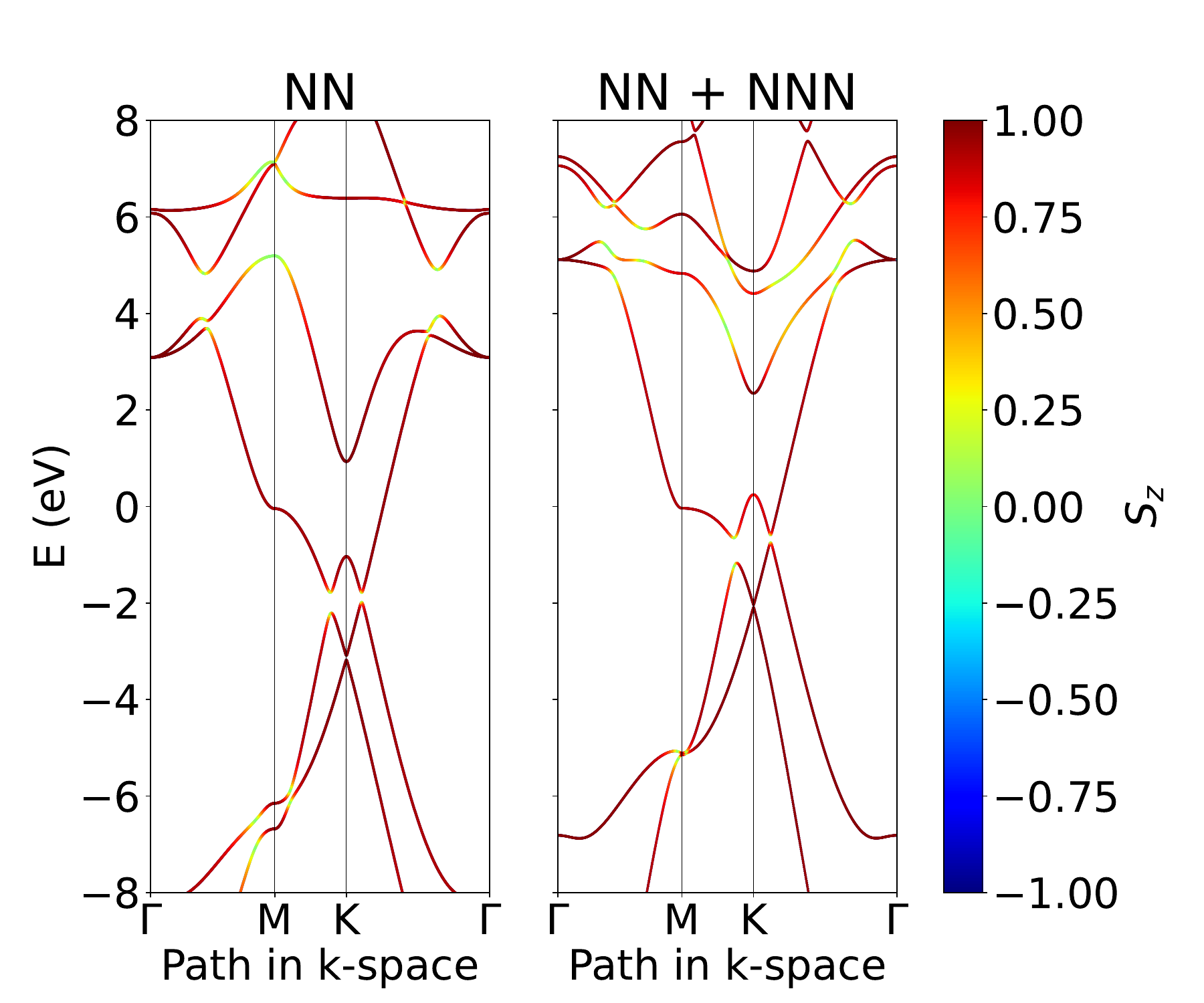} 
		\caption{}
        \label{fig:only_SOC_kagome}
    \end{subfigure}
 \caption{The band structure of the kagome lattice in an interval between $-8 \ \si{\eV}$ and $8 \ \si{\eV}$. The bands are color-coded according to the expectation value $\langle S_z \rangle$. Different interactions are discussed:
    (a) only Zeeman interaction ($\lambda_{\text{ZM}} = 0.2 \ \si{\eV}$) (b) both Zeeman ($\lambda_{\text{ZM}} = 0.2 \ \si{\eV}$) and SOC ($\lambda_{\text{SOC}} = 1.0 \ \si{\eV}$) (c) only SOC ($\lambda_{\text{SOC}} = 1.0 \ \si{\eV}$). In the latter case, the bands remain fully degenerate because time-reversal and inversion symmetry are preserved. Only the band with $\langle S_z \rangle \approx 1$ is visible.}
\end{figure}

Subsequently, we want to discuss the effect of SOC on the electronic band structure, as well as the spin polarization, in detail. 
The kagome lattice still serves as the basic reference system for the following discussion. To better understand the effect of SOC, one would start by analyzing a Zeeman interaction, since its effect on the band structure is much more straightforward.
The Zeeman interaction energetically favors the spin-up orbitals by lowering their on-site energy by an amount of $\lambda_{\ZM}$, while raising the on-site energy of spin-down orbitals by the same amount. This leads to a uniformly splitting of spin-up and spin-down. This term also
commutes with $S_z$ and thus preserves $S_z$. Hence, when only the Zeeman interaction is present, the spin polarization $S_z$ remains a good quantum number. In Fig.~\ref{fig:only_Zeeman_kagome}, the calculated expectation value $\langle S_z \rangle$ along a plot of the band structure is shown. In that case, $\langle S_z \rangle$ only takes value of either $1$ or $-1$ (in the unit of $\hbar/2$), which suits the expectation. 

In contrast to the widely used SOC term proposed by Kane and Mele \cite{KaneMelePRL2005}, the atomic SOC term in our model does not commute with $S_z$ and allows spin mixing processes \cite{JAFFE1987399}. Therefore, in the presence of this term, $S_z$ is not preserved and thus not a well-defined quantum number. In Fig.~\ref{fig:Zeeman_SOC_kagome}, both Zeeman interaction and SOC are included in the calculation. Under the influence of SOC, $\langle S_z \rangle$ now allows for continuous values between $-1$ and $1$. We can see that some crossing points between spin-up and
spin-down bands are lifted and bands with new characters are formed: the spin changes its sign (from up to down and vice versa) 
along these bands. Since the spin changes sign, there must be regions on the band where the value of $\langle S_z \rangle$ approaches $0$, 
i.e. the spin turns into the $xy$-plane.
The points at which $\langle S_z \rangle = 0$ are commonly noted as avoided crossing and often go along with lifted degeneracies. In Fig.~\ref{fig:Zeeman_SOC_kagome},
they can be recognized by the light green color corresponding to $S_z = 0$.

\begin{figure}[H]
    \begin{subfigure}{0.32\linewidth}
        \centering
        \includegraphics[width=\textwidth]{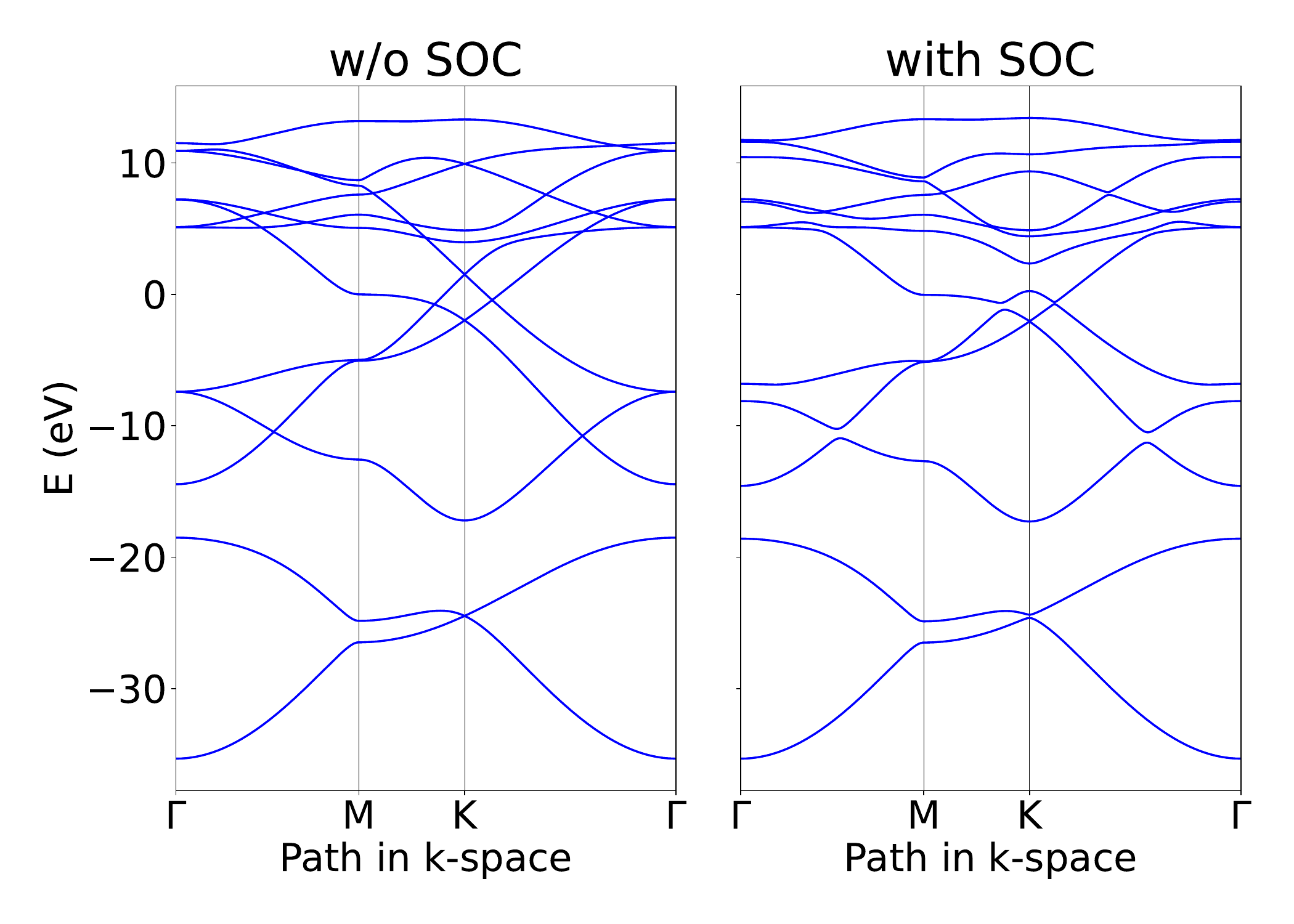} 
		\caption{}
        \label{fig:kagome_full_BS}
    \end{subfigure}
    \begin{subfigure}{0.32\linewidth}
        \centering
        \includegraphics[width=\textwidth]{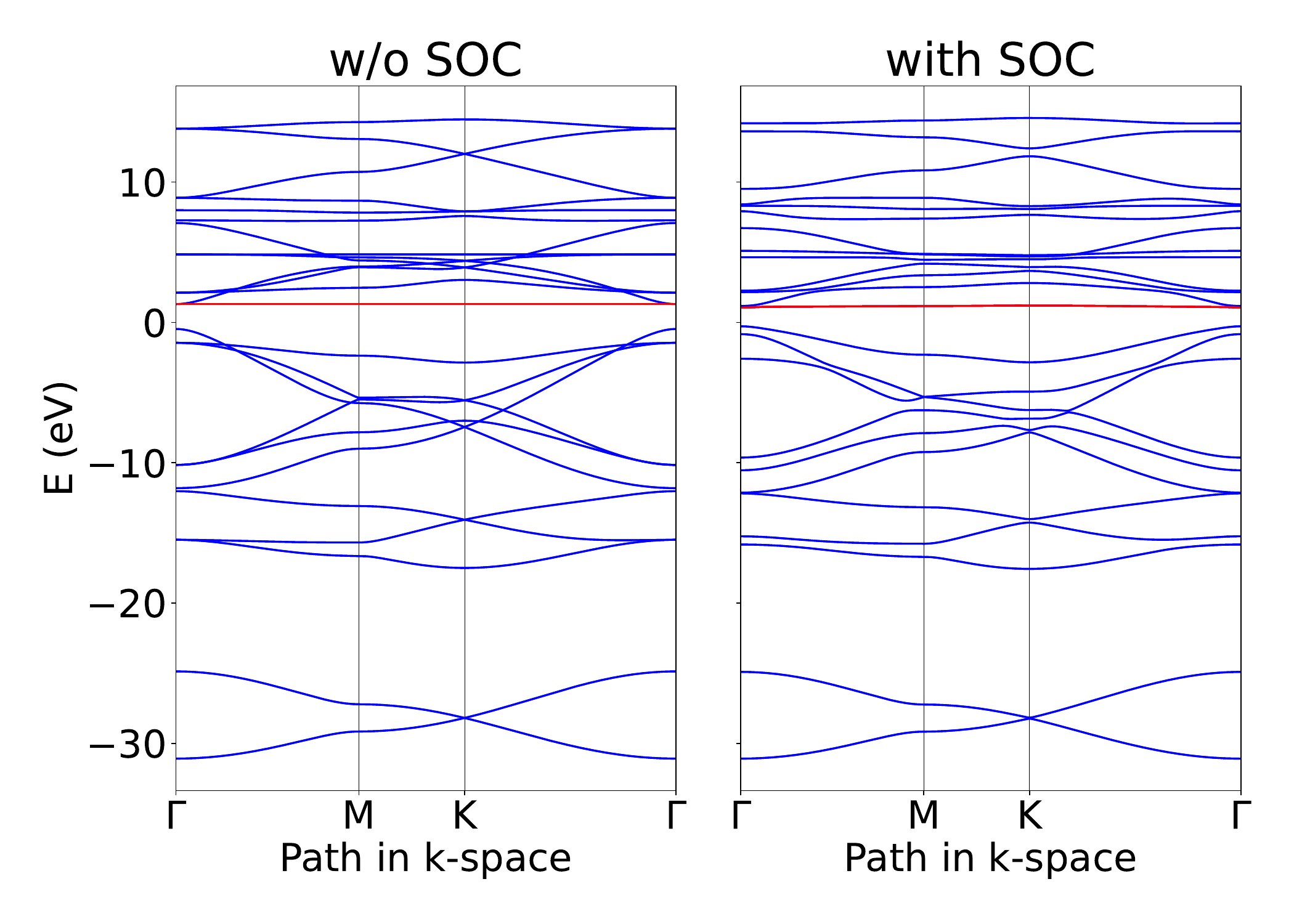} 
		\caption{}
        \label{fig:truncated_hex_full_BS}
    \end{subfigure}
    \begin{subfigure}{0.32\linewidth}
        \centering
        \includegraphics[width=\textwidth]{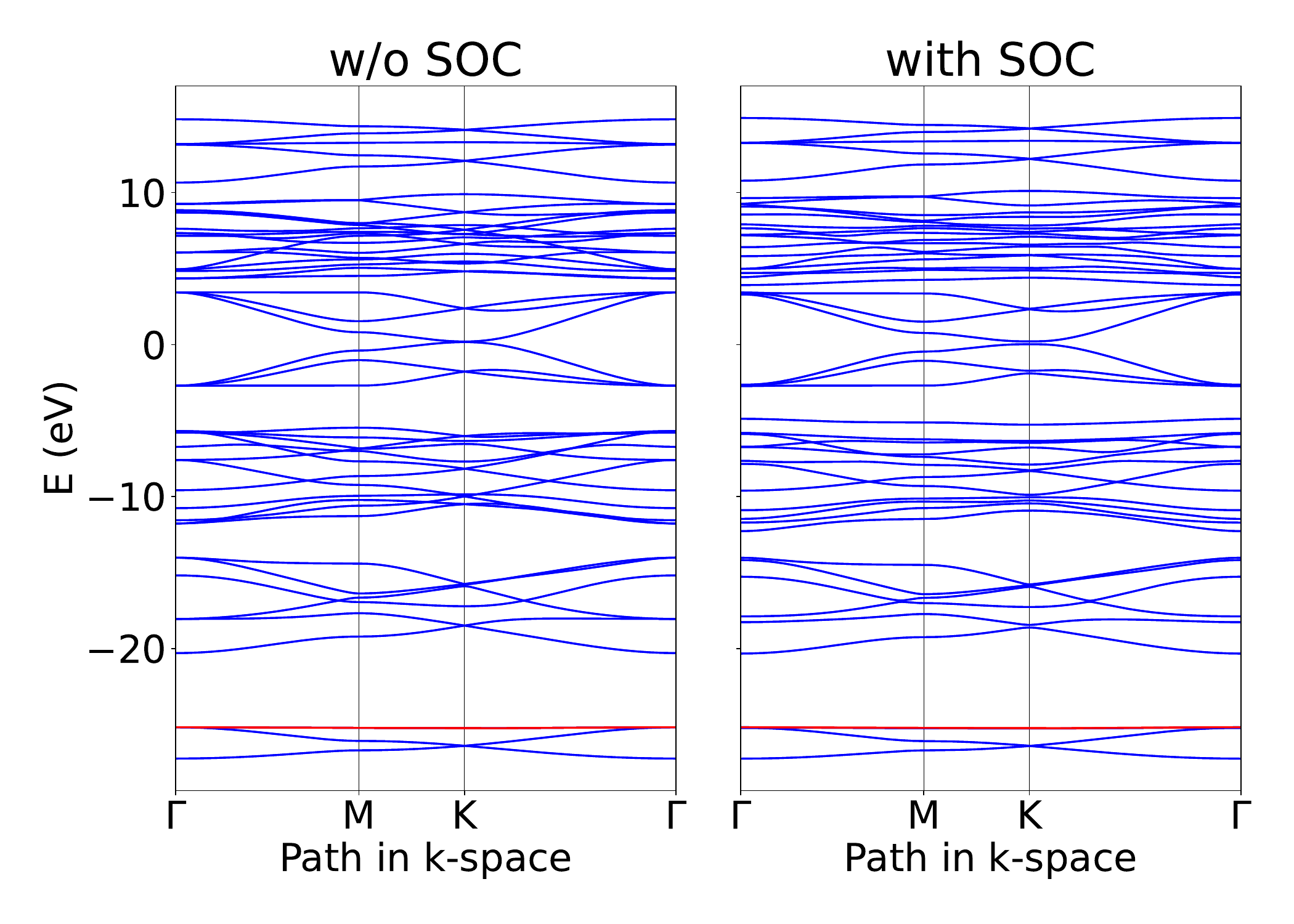} 
		\caption{}
        \label{fig:truncated_trihex_full_BS}
    \end{subfigure}
    \caption{Band structures of (a) the kagome (b) the truncated hexagonal and (c) the truncated trihexagonal lattices with and without SOC. Both NN and NNN hopping terms are included. The strength of SOC, if present, is $1 \ \si{\eV}$.}
    \label{fig:kagome_truncated_hex_trihex_bandstruct}
\end{figure}

In Fig.~\ref{fig:only_SOC_kagome}, where only SOC is present,these spin-mixing effects are prominently present, highlighted in green, i.e. $\langle S_z \rangle \approx 0$, at many points on multiple bands. In contrast to isolated atomic systems, where SOC always splits the energy levels of up and down spin, in solids this degeneracy can be protected by crystal symmetries \cite{dresselhaus2007group}. By comparing Fig.~\ref{fig:Zeeman_SOC_kagome} and Fig.~\ref{fig:only_SOC_kagome} to Fig.~\ref{fig:only_Zeeman_kagome} in calculations where NNN hopping terms are not included, it is also evident that the well known flat band of the kagome lattice gains dispersion when SOC is present. In Fig.~\ref{fig:kagome_truncated_hex_trihex_bandstruct}, the full band structures of the kagome, the truncated hexagonal the truncated trihexagonal lattices are shown. The truncated hexagonal and truncated trihexagonal also display rich electronic properties with multiple fully flat bands and Dirac/high degeneracy points. However, most of these points are lifted in the presence of SOC. Interestingly, some flat bands of these two lattices still remain flat over the whole Brillouin zone even with NNN hopping terms and SOC present. In Fig.~\ref{fig:kagome_truncated_hex_trihex_bandstruct}, these bands are highlighted in red. Since the group velocity is proportional to $\partial E/ \partial \kk$, states on these nearly flat bands are highly localized in real space and exhibit a very narrow bandwidth. As a result, the Coulomb interaction energy between electrons in these states exceeds their kinetic energy, leading to strongly correlated behavior. Consequently, the Archimedean lattices presented here may serve as promising platforms for realizing and exploring such correlated electronic phases.

We want to remind, that we restrict our study to perfectly planar carbon networks, as this assumption captures the essential electronic features of the Archimedean lattices. While buckling may occur in realistic 2D systems due to substrate interactions or strain, such structural deviations are not expected to qualitatively alter the topological characteristics of the bands. In fact, moderate buckling in $sp^2$-based materials is known to further flatten bands and increase their energetic separation~\cite{PhysRevB.102.245427,vanpoppelen2025bucklingflatbandstwisted,PhysRevB.111.235407,Lopez-Suarez_buckling}, which could even favour spin-orbit coupling induced band inversions. Thus, our approach should be regarded as a proof-of-principle description of the intrinsic electronic and topological properties of planar Archimedean lattices.

All band structures of the remaining Archimedean lattices are provided in the Supplementary Information~\cite{SI} in the same manner. 

\subsection{Ribbon configurations \& edge states}

\begin{figure}[H]
    \begin{subfigure}{\linewidth}
        \centering
        \includegraphics[width=0.99\linewidth]{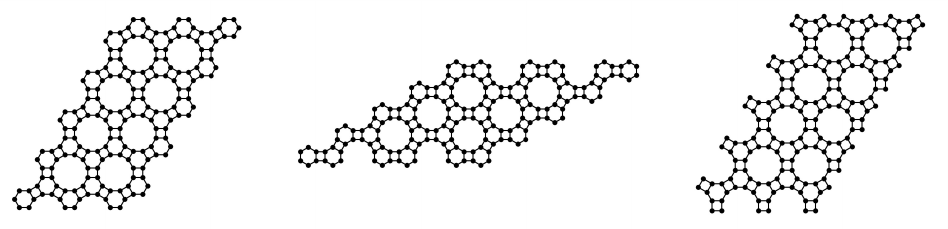} 
		\caption{}
        \label{fig:truncated_trihexagonal_ribbon}
    \end{subfigure}
    \begin{subfigure}{\linewidth}
        \centering
        \includegraphics[width=0.99\linewidth]{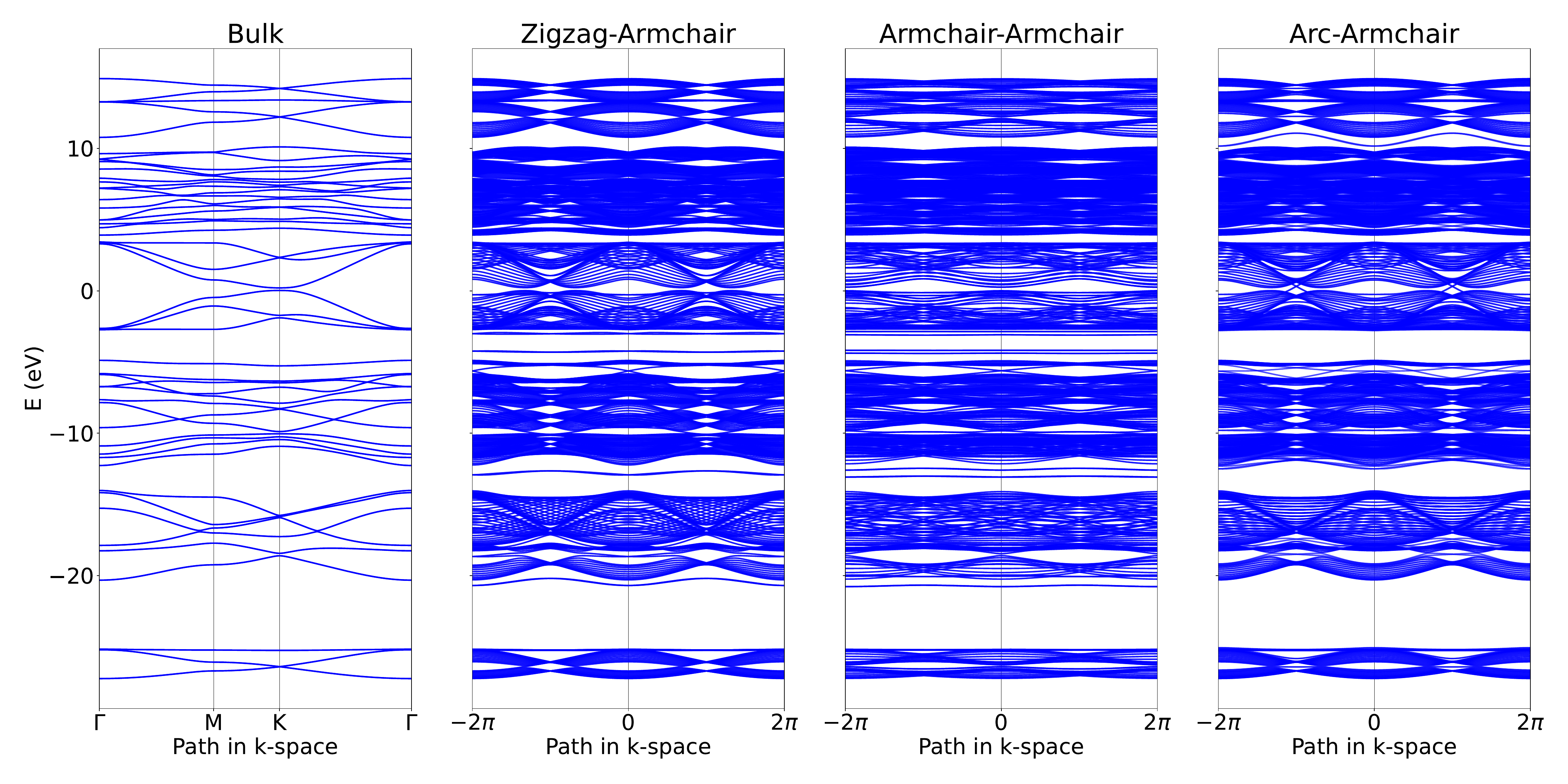} 
		\caption{}
        \label{fig:truncated_trihexagonal_ribbon_bandstructure}
    \end{subfigure}
   \caption{(a) From left to right: zigzag-armchair, armchair-armchair, arc armchair ribbons of the truncated trihexagonal lattice (b) Bulk and ribbon band structures of the truncated trihexagonal lattice. Both NN and NNN hopping terms, as well as SOC with $\lambda_{\text{SOC}} = 1 \ \si{\eV}$ are included. The periodicity is set in the horizontal direction for all ribbons. Here, all band structure calculations are performed on ribbons with a width of 12 unit cells in the non periodic direction and 1 unit cell in the periodic direction.}
\end{figure}

Based on the method described in Sec.~\ref{sec:method_ase}, nanoribbon structures can be generated in a straightforward manner. By selecting different lattice vectors and atomic sites within the unit cell, various edge configurations can be constructed. Here, we present three distinct edge terminations of the truncated trihexagonal lattice—the zigzag–armchair, armchair–armchair, and arc–arm\-chair edges (Fig.~\ref{fig:truncated_trihexagonal_ribbon})—together with their corresponding band structures (Fig.~\ref{fig:truncated_trihexagonal_ribbon_bandstructure}). To the best of our knowledge, the latter configuration has not been discussed previously. 
We find several types of new states emerging within the bulk continuum. Fully isolated, weakly dispersive bands appear in continuum holes, characteristic of Tamm- or Shockley-type edge states that are strongly localized and sensitive to edge termination, e.g., in the band gaps around $-4$ and $-12~\si{\eV}$, visible for the zigzag–armchair and armchair–armchair ribbons but absent for the arc–armchair edges. Other bands exhibit pronounced hybridization with bulk states, forming edge resonances, while some display continuous crossings through small bulk gaps, indicative of probable topological edge modes connecting valence and conduction bands, which will be discussed in the following section.
Topological edge states in two-dimensional materials can be experimentally probed using scanning tunneling microscopy and spectroscopy (STM/STS) to map their spatial localization, or by angle-resolved photoemission spectroscopy (ARPES) to resolve their energy–momentum dispersion~\cite{WuARPESedge,TangARPES,UgedaSTM}.

\section{Hall effects \& topological properties}
The characterization of the quantum spin Hall (QSH) phase requires the evaluation of the $\mathbb{Z}_2$ topological invariant. There are numerous methods to calculate this invariant such as
the Wannier charge centers approach (as proposed by Soluyanov and Vanderbilt \cite{soluyanov2011computing}) or time-reversal polarisation (proposed by Fu and Kane  \cite{PhysRevB.74.195312}). In the latter method, one has to define a matrix $w_{mn} = \langle u_m(-\kk) | \mathcal{T} | u_n (\kk) \rangle$ from the Bloch functions $| u_n (\kk) \rangle$ of occupied states and the time-reversal symmetry operator $\mathcal{T}$. On the high-symmetry points $\Lambda_i$ in the Brillouin zone, or the so-called \emph{time-reversal momenta} (TRIM), $w_{mn}$ is antisymmetric. Consequently, one can calculate the quantity $\delta_i = \Pf\!\big[w(\Lambda_i)\big]\Big/\sqrt{\det\!\big[w(\Lambda_i)\big]} = \pm 1$ on all TRIM point. In the 2D case, there are in total 4 TRIM points in the Brillouin zone, so the $\mathbb{Z}_2$ invariant $\nu$ is given as $(-1)^\nu = \prod_{i = 1}^4 \delta_i$. In a follow-up work, Fu and Kane showed that this method can be greatly simplified when the system has inversion symmetry \cite{fu2007topological}. At the TRIM points, Bloch states $| u_n (\Lambda_i) \rangle$ are also parity eigenfunctions with eigenvalues $\xi_n(\Lambda_i) = \pm 1$. The quantity $\delta_i$ simplifies to $\delta_i = \prod_n \xi_n (\Lambda_i)$ where n runs over $N_F/2$ Kramers pairs of $N_F$ occupied bands. The $\mathbb{Z}_2$ invariant can again be calculated by taking the product of $\delta_i$ over all TRIM points:
\begin{equation}
  (-1)^{\nu}
  = \prod_{i=1}^{4} \delta_i
  = \prod_{i=1}^{4} \prod_{n=1}^{N_F/2} \xi_n(\Lambda_i)\,.
  \label{eq:Z2_invariant}
\end{equation}
Since all Archimedean lattices have inversion symmetry,the time-reversal polarisation method was applied to characterize the QSH phase. Nevertheless, we note that the $\mathbb{Z}_2$ invariant is generally regarded as the defining topological index of the quantum spin Hall phase, which is typically equivalent to the topological insulator phase in time-reversal symmetric systems. However, special cases have been reported in which a topological phase characterized by a finite spin Chern number remains $\mathbb{Z}_2$-trivial~\cite{Wang_2024}.

Consider a nanoribbon setup, where the system is in the QSH phase and shares at least one interface with a topologically trivial material, e.g. vacuum. According to the bulk–boundary correspondence, there should be pairs of edge states bridging the band gap \cite{hasan2010colloquium}. These edge states are topologically protected by time-reversal symmetry, which forbids elastic backscattering between counter-propagating states of opposite spin. As a result, they remain robust against non-magnetic disorder and structural imperfections, preserving dissipationless edge transport characteristic of the quantum spin Hall phase. 
Hence, we search for topological edge states in the ribbon band structures within the same energy window that corresponds to the occupation level at which the QSH phase occurs, i.e., where the product in Eq.~(\ref{eq:Z2_invariant}) equals $-1$. In this regime, a band gap should be present in the ribbon band structure, and pairs of topological edge states must connect the valence and conduction bands by bridging this gap.

\begin{figure}[H]
    \centering
    \includegraphics[width=0.99\linewidth]{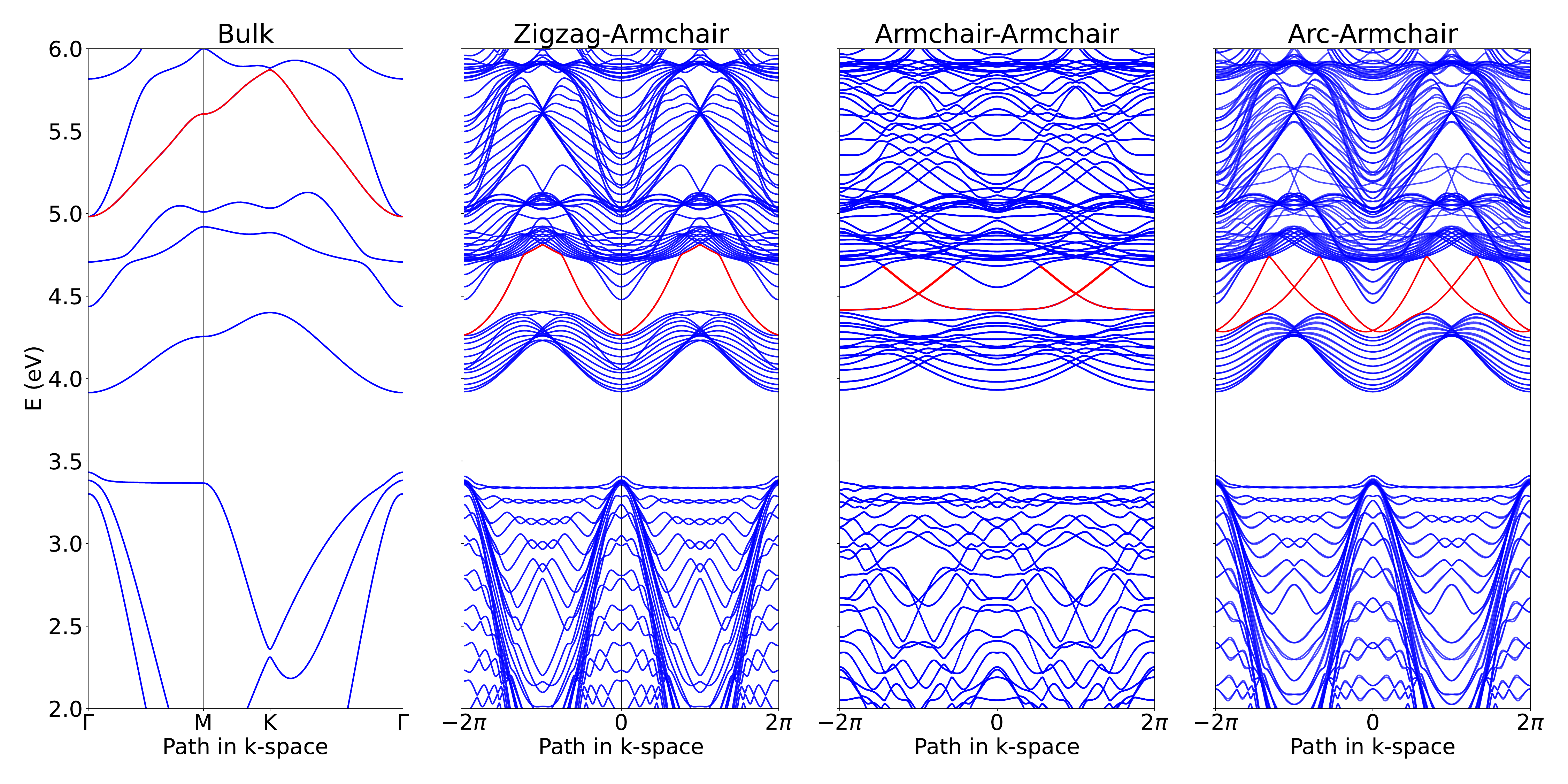}
    \caption{Bulk and ribbon band structures of the truncated trihexagonal lattice. In the ribbon band structures, the topological edge states are highlighted in red. The red band in the bulk band structure indicates the change in the $\mathbb{Z}_2$ topological index $\nu = 0 \rightarrow 1$. The Fermi energy is $E_F \approx 0.7 \si{\eV}$.}
    \label{fig:truncated_trihex_topological}
\end{figure}

Fig.~\ref{fig:truncated_trihex_topological} presents the bulk band structure of the truncated trihexagonal lattice together with the ribbon band structures of three different edge configurations as shown in Fig.~\ref{fig:truncated_trihexagonal_ribbon}. The band marked in red in the bulk band structure is the occupation at which the system displays the QSH phase, i.e. $\mathbb{Z}_2$ index $\nu=1$. In the same energy window there is a fundamental band gap between $4.4 \ \si{\eV}$ and $4.5 \ \si{\eV}$ covering the whole Brillouin zone. In all ribbon band  structures, we found two pairs of helical, topological edge states bridging this global gap, as expected by the bulk–boundary correspondence. The edge states are highlighted in red in Fig.~\ref{fig:truncated_trihex_topological}. In our tight-binding calculations, the topologically protected edge states appear a few $\si{\eV}$ away from the Fermi energy. While conventional electrostatic gating through SiO$_2$ substrates typically allows Fermi-level tuning on the order of a few hundred meV, larger shifts of up to $\sim 1\,\si{\eV}$ and beyond have been experimentally achieved by ionic-liquid gating in graphene and related 2D materials~\cite{WangGating,KakenovGating,ChuGating}. We note that the exact position of the Fermi energy is not the focus of our present proof-of-principle tight-binding model and may be modified in realistic implementations, for example, by charge transfer from substrates, chemical doping, or external gating.

\begin{figure}[H]
    \begin{subfigure}{0.32\linewidth}
        \centering
        \includegraphics[width=\linewidth]{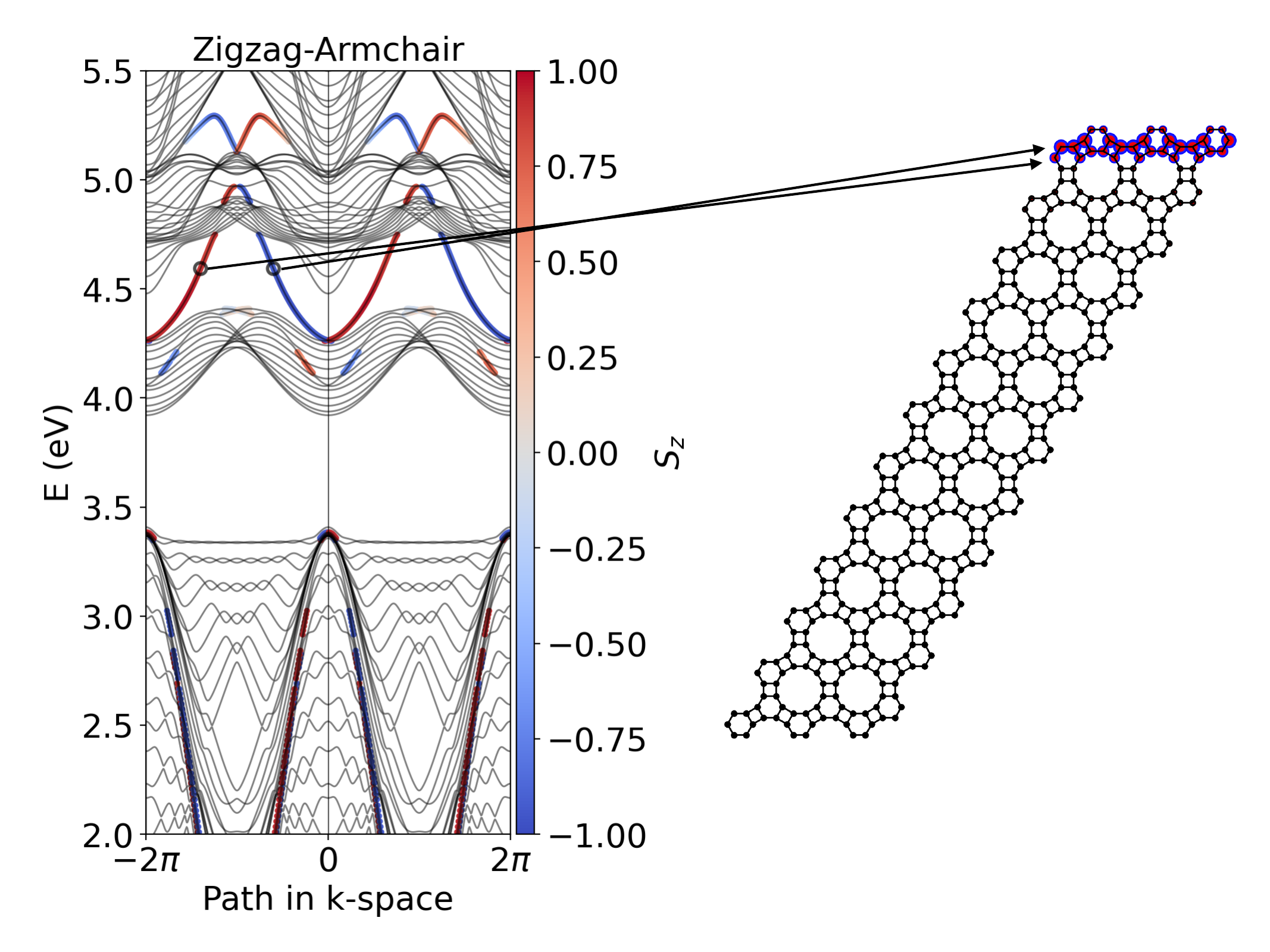} 
		\caption{}
        \label{fig:truncated_trihex_zigzag_spin}
    \end{subfigure}
    \begin{subfigure}{0.32\linewidth}
        \centering
        \includegraphics[width=\linewidth]{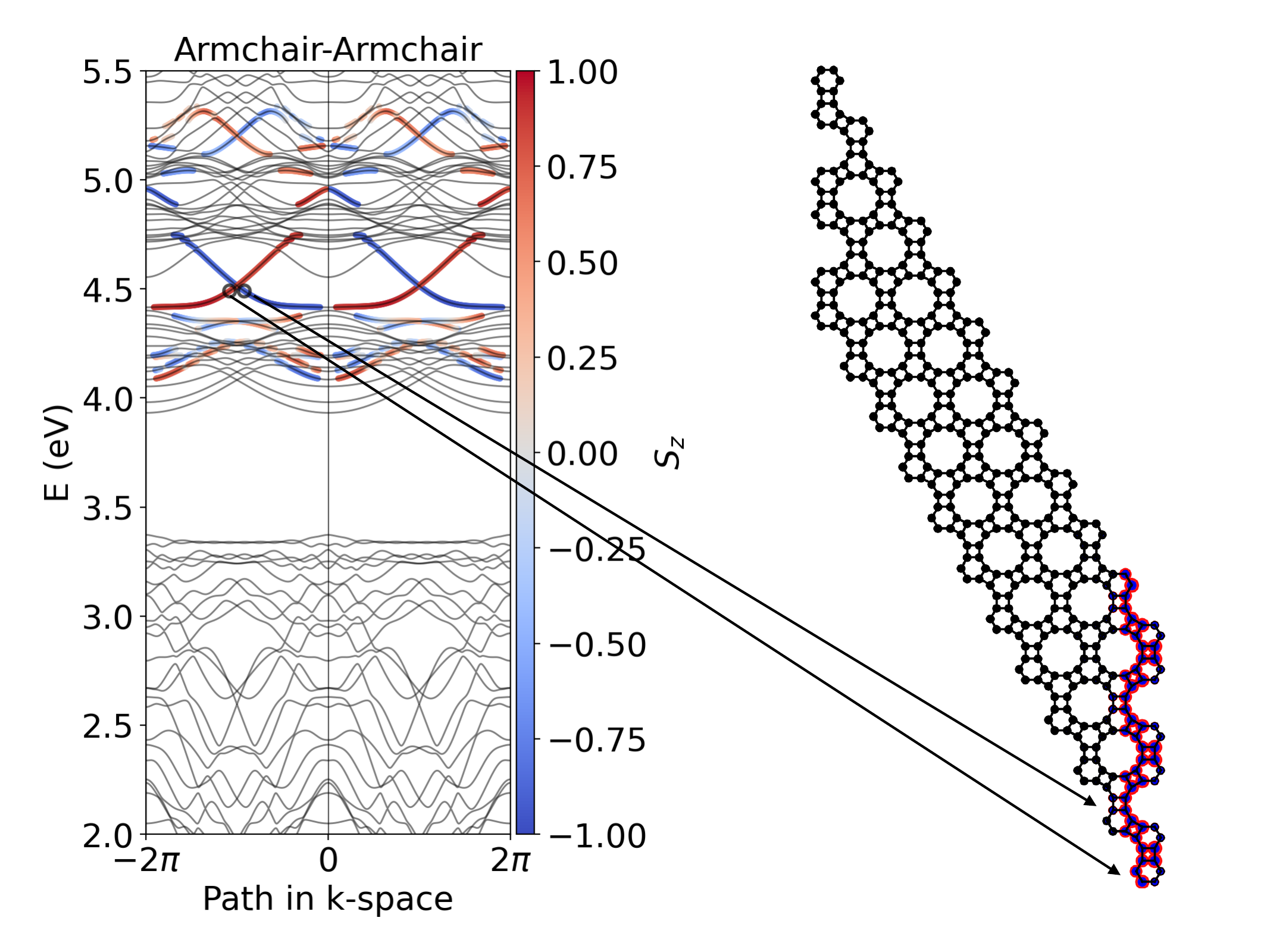} 
		\caption{}
        \label{fig:truncated_trihex_armchair_spin}
    \end{subfigure}
    \begin{subfigure}{0.32\linewidth}
        \centering
        \includegraphics[width=\linewidth]{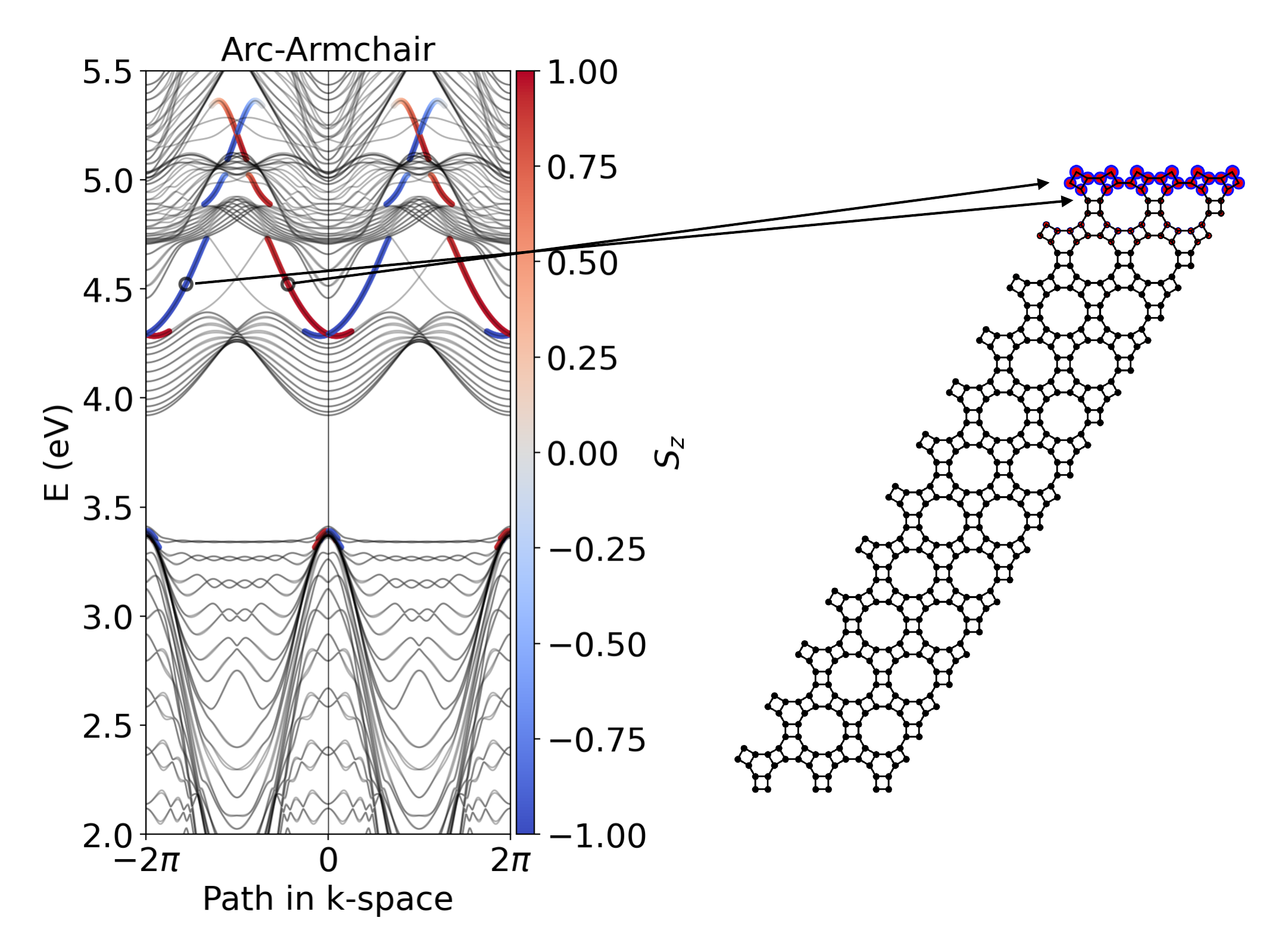} 
		\caption{}
        \label{fig:truncated_trihex_arc_spin}
    \end{subfigure}
    \caption{The band structure of the truncated trihexagonal (a) zigzag-armchair (b) armchair-armchair (c) arc armchair ribbons. Band structure calculations are performed on ribbons with a width of 12 unit cells in the non periodic direction and 1 unit cell in the periodic direction. We projected the edge states according to their exponentially decaying profiles and color-coded them by the value of $\langle S_z \rangle$. The representative states are visualized based on their spatial distributions: the point size reflects the probability density, while the color indicates the dominant spin polarization. With time-reversal and inversion symmetry preserved, the sign of $\langle S_z \rangle$ is reversed on the opposite edge. In the special case of the arc armchair ribbon, the edges are dissimilar, and the edge states originating from the opposite edge can also be seen crossing the gap as non-highlighted bands. The calculations were done for ribbons with only one unit cell in the periodic direction, but the results are periodically replicated for a better visualization on ribbons with a larger width of 3 unit cells in this direction.}
    \label{fig:truncated_trihex_ribbons_spin}
\end{figure}

A spin polarization analysis (Fig.~\ref{fig:truncated_trihex_ribbons_spin}) reveals that the states found in our calculation have a high degree of spin polarisation. To demonstrate this, a pair of points on the spin-up and spin-down polarized topological edge states is chosen for each ribbon band structure (these point of marked by black circles in Fig.~\ref{fig:truncated_trihex_ribbons_spin}). All chosen points have $\left|\langle S_z\rangle\right| > 0.8$
(see the caption of Fig.~\ref{fig:truncated_trihex_ribbons_spin} for exact values).
The selected states are also visualized through their spatial distributions to emphasize their characteristic localization at the edges.
In contrast to the SOC model used by Kane and Mele, where there is no interaction between spin-up and spin-down states,
the atomic SOC used in our model enables this interaction and thus allows for spin-flip processes. For this reason, $\langle S_z \rangle$ is not exactly $\pm 1$. However, since the spin polarization of these edge states is still very high, the probability for the occurrence of spin-flip processes is very small \cite{fabian99,gradhand2010}. Therefore, the topological edge states found in our calculations still retain their
special feature: they are robust spin current channels with a high degree of spin polarisation allowing for almost dissipationless spin transport. The latter renders them potentially relevant for applications in the field of spintronics. 

With the confirmed existence of topological edge states enabling stable spin-current transport, it becomes meaningful to evaluate the intrinsic spin Hall conductivity. This quantity can be determined entirely from the bulk band structure of the system under consideration. Similar to the quantum anomalous Hall effect, where the intrinsic Hall conductivity
can be described by the Berry curvature, the intrinsic spin Hall conductivity can be expressed as a function of the Fermi energy stressing the
Kubo formula approach \cite{RevModPhys.82.1539, RevModPhys.87.1213, RevModPhys.82.1959, PhysRevB.104.184423}:
\begin{equation}
    \sigma_{xy}^z (E_F) = \frac{e}{\hbar} \sum_{n} \frac{1}{(2 \pi)^2} \int_{E(\kk) \leq E_F} \Omega^z_{xy, n} (\kk) d^2 \kk \,.
\end{equation}
The integration on the whole Brillouin zone is performed over all occupied states in the band structure
for this quantity. The Fermi energy fixes the filling of the bands and in practice can be varied by applying an external bias voltage or chemical doping. 
Here $\Omega_{xy, n}^z (\kk)$ is the \emph{spin Berry curvature}
of the $n$-th band, defined as 
\begin{equation}
    \Omega_{xy, n}^z (\kk) = -2 \hbar^2 \Imag \sum_{m \neq n} 
    \frac{ \langle n, \mathbf{k}| \Sigma^z_x | m, \mathbf{k} \rangle \langle m, \mathbf{k}| v_y | n, \mathbf{k} \rangle}{[E_{n}(\mathbf{k} )- E_{m}(\mathbf{k})]^2} \,,
    \label{eq:spin_Berry_curvature}
\end{equation}
where $|n, \kk \rangle = \varphi_n (\kk)$ and $E_n (\kk)$ are the eigenvectors and
the eigenenergies obtained by means of the tight-binding model, respectively. $v_y$ is the partial derivative of the Hamiltonian in the $k_y$ direction and $\Sigma^z_x$ is the spin-current operator, defined as the anticommutator of the velocity operator $v_x$ and the Pauli matrix $s^z$: $\Sigma^z_x = \frac{1}{2} \{ v_x, s^z \}$. In our calculations, the first derivatives of the Hamiltonian are calculated numerically with a finite difference method, using grids of size $101 \times 101$ points and third order of accuracy, i.e. 6 neighbors points are used in the calculation for each point. 

\begin{figure}[H]
    \centering
    \includegraphics[width=0.99\linewidth]{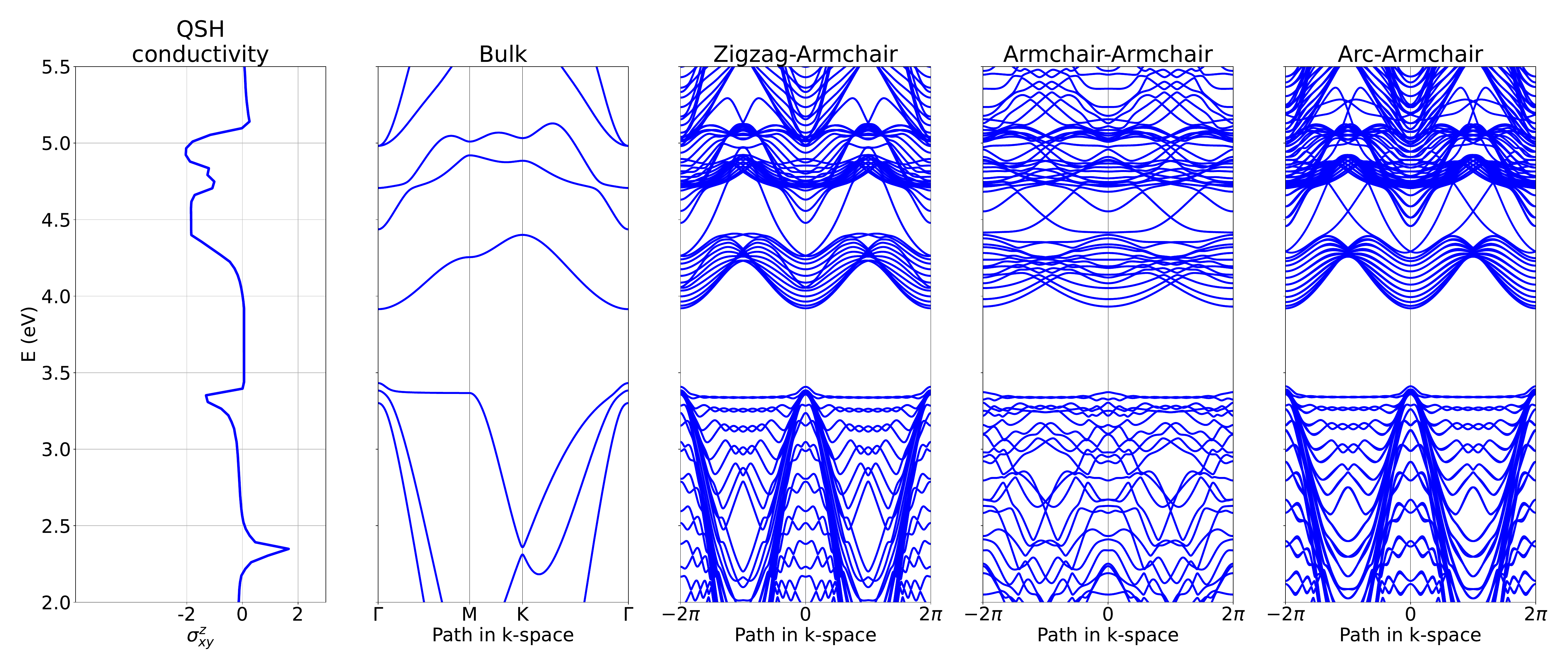}
    \caption{The intrinsic spin Hall conductivity of the truncated trihexagonal lattice together with its bulk and ribbon band structures in the energy interval where topological edge states were found. Inside the topologically nontrivial band gap, the spin Hall conductivity exhibits large, quantized values ($\sigma_{xy}^z (E \approx 4.5 \si{\eV}) \approx e/\pi$), a hallmark of the quantum spin Hall phase.}
    \label{fig:truncated_trihex_QSH_topological}
\end{figure}

Figure~\ref{fig:truncated_trihex_QSH_topological} presents the intrinsic spin Hall conductivity of the truncated trihexagonal lattice (in units of $e/2\pi$), along with its bulk and ribbon band structures, in the energy range corresponding to the topological edge states identified previously. 

Experimentally the quantum spin Hall effect can be identified via quantized edge conductance and corroborated by ARPES or spin-resolved STM/\-STS imaging of spin-polarized edge bands; by contrast, the spin Hall effect is detected through spin accumulation at sample edges, measured with optical Kerr/Faraday rotation or spin-torque methods \cite{KatoSHE,Valenzuela_2006,Buhrman_spintorque}.

While the QSH phase can be understood as two time-reversed Chern insulators with opposite Chern numbers $C_\uparrow = -C_\downarrow$, the associated spin Chern number $C_s = \frac{1}{2}(C_\uparrow - C_\downarrow)$ quantizes the spin Hall conductivity as $\sigma_{xy}^s = \frac{e}{2\pi} C_s$ only if $s_z$ is (approximately) conserved. In realistic systems with strong spin–orbit coupling, spin is generally not a good quantum number (c.f. Fig.~\ref{fig:truncated_trihex_ribbons_spin}), and inter-spin mixing destroys exact quantization, thus leaving only the parity of $C_s$, the aforementioned $\mathbb{Z}_2$ index $\nu$, as a topologically protected quantity. Consequently, the QSH conductance remains robust against disorder and perturbations but is not strictly quantized in spin units.

Inside the topologically nontrivial, inverted band gap at $E \approx 4.5 \si{\eV}$ the spin Hall conductivity exhibits these almost perfectly quantized values of $\sigma_{xy}^z \approx e/\pi$--a hallmark of the quantum spin Hall phase. The aforementioned topological edge states, which serve as robust spin-current channels protected by time-reversal symmetry, are precisely the states responsible for this non-zero conductivity. As can be seen from the ribbon band structures in Figs.~\ref{fig:truncated_trihex_arc_spin} and \ref{fig:truncated_trihex_QSH_topological} the truncated trihexagonal lattice hosts two Kramers pairs of helical edge states within the non-trivial bulk gap (i.e., two spin-filtered channels per edge). Hence, the spin Chern number is 
$ C_s = 2 $, and the corresponding quantized spin Hall conductivity becomes $\sigma_{xy}^{s} \approx \frac{e}{2\pi} C_s \approx \frac{e}{\pi}$. 

Expectedly, the conductivity in the global trivial band gap at around $3.5 \allowbreak \dots\allowbreak 4.0\ \si{\eV}$ is zero, since there are no topological edge states bridging this gap. 
Even in globally trivial systems ($\nu = 0$), finite spin Hall conductivities may arise from intrinsic Berry curvature near avoided crossings when the Fermi level lies within overlapping bands. However, if the Fermi energy resides inside a \emph{global trivial gap}, the bulk contribution vanishes, as the integrated spin Berry curvature of all occupied states cancels exactly. In this case, any apparent, locally quantized plateaus in $\sigma^{z}_{xy}(E)$ originate not from bulk topology but from quasi-bound, spin-polarized edge or interface states that transiently carry spin current along the sample boundaries within the gap. The underlying bulk thus strictly remains topologically trivial with zero intrinsic spin Hall conductivity.

The intrinsic spin Hall conductivities of the remaining seven Archimedean lattices, together with a detailed discussion of the truncated hexagonal lattice, are provided in the Supplementary Information~\cite{SI}. While the truncated hexagonal lattice hosts a quantum spin Hall phase far from a realistic Fermi level, it exhibits strikingly edge-dependent behavior, featuring strongly localized, resonance-like states weakly hybridized with the bulk on one edge and heavily hybridized, bulk-like bands on the other.

\section{Conclusion \& outlook}

In the presented work, we developed a Python-based tight-binding framework for two-dimensional carbon networks that enables the construction of Archimedean lattice geometries in both bulk and ribbon configurations. The model incorporates $s$ and $p$ orbitals and an atomistic description of spin–orbit coupling (SOC), allowing for a unified investigation of electronic structures and spin-dependent transport phenomena. Using this framework, we systematically analyzed the topological and spin Hall properties across all eight pure Archimedean lattices, distinguishing between genuinely topological quantum spin Hall (QSH) phases and trivial systems exhibiting spin-resolved resonance states.

Several lattices, such as the truncated hexagonal and the truncated trihexagonal lattices, were found to host fully flat bands extending throughout the Brillouin zone. Remarkably, these bands remain nearly dispersionless even when next-nearest-neighbour (NNN) hopping and a large SOC strength of $\lambda_{\mathrm{SOC}} = 1 \ \si{\eV}$ are included. Such robustness makes these systems promising candidates for the study of strongly correlated electron behavior, where flat bands can amplify interaction effects and drive emergent quantum phases. We furthermore identified Dirac fermions and high-degeneracy band-crossing points in the snub-trihexagonal, truncated-trihexagonal, rhombitrihexagonal, and kagome lattices--most likely arising from symmetry-protected band crossings inherent to their underlying triangular network geometry. These symmetry-enforced degeneracies are of interest, as they give rise to massless Dirac quasiparticles, linear band dispersions, and provide a natural platform for realizing topological and spin-orbit–driven phases. 

Among all investigated geometries, the truncated trihexagonal lattice stands out for its rich topological character. Multiple ribbon terminations of this lattice were found to host helical edge states that persist across different edge configurations, underlining their topological protection. These states carry a quantized spin Hall current, confirming their role as robust spin current channels. The coexistence of multiple Kramers pairs within the bulk gap further enhances the spin Hall response, making this lattice particularly appealing for spintronic applications.  

In contrast, the other Archimedean lattices either exhibit trivial band topology with occasional resonance-like edge states appearing inside local bulk gaps, or display nontrivial band inversions that, however, lack a global band gap and thus remain semi-metallic, preventing a quantized spin Hall conductivity. In some cases, such as the truncated hexagonal lattice, nontrivial phases with a global band gap are found only at energies far from a realistic Fermi level \cite{SI}.

We propose the future work will extend these studies beyond 2D carbon allotropes toward systems composed of heavier elements with intrinsically strong SOC that might form these lattices. Evaluating Archimedean lattices based on transition-metal dichalcogenides (TMDCs) or other high-SOC materials through high-throughput or machine-learning-based approaches will be a promising direction \cite{jonathanschmidtML, Haichenacs.jpclett.3c02131, wang2023symmetry, Haastrup_2018, Gjerding_2021}. Such investigations may reveal new topological regimes or enhance spin-dependent transport responses without the necessity for proximity-induced SOC.

In addition, the recently proposed orbital Hall effect (OHE) \cite{PhysRevLett.121.086602} has opened an exciting avenue in the field of orbitronics. A first study on simple $s$-orbital-based Archimedean lattices without SOC has shown that both geometry and electronic structure crucially determine the emergence of an orbital Hall current \cite{OliverBuschOH}. Building on this, the exploration of the OHE in more complex Archimedean geometries--particularly the truncated hexagonal and truncated trihexagonal lattices--will provide deeper insight into the interplay of orbital motion, lattice symmetry, and topology. Such studies are currently in preparation and will be reported in forthcoming work by the authors.

\section*{Data availibilty statement}
All data necessary to reproduce the results presented in this work are available from the authors via an open data repository~\cite{zenodo_pham_hinsche_Mertig_2025}. The structural data of all eight pure Archimedean lattices are provided as Crystallographic Information Files (CIFs). The full Python code used in this study is openly available and includes minimal working examples to compute bulk and ribbon band structures, as well as derived quantities such as spin polarizations, $\mathbb{Z}_2$ invariants, and spin Hall conductivities. Additional data supporting the findings of this study are available from the authors upon reasonable request.

\section*{Acknowledgment}
We would like to thank Miguel A. L. Marques for encouraging us to explore the intriguing geometrical properties of Archimedean lattices. We would like to thank Franziska Ziolkowski and Robin R.~Neumann for enlightening discussions. LVDP gratefully acknowledges financial support by a stipend of the Collaborative Research Centre 1415 ''Chemistry of Synthetic Two-Dimensional Materials''.

\printbibliography

@article{OliverBuschOH,
  title = {Orbital Hall effect and orbital edge states caused by $s$ electrons},
  author = {Busch, Oliver and Mertig, Ingrid and G\"obel, B\"orge},
  journal = {Phys. Rev. Res.},
  volume = {5},
  issue = {4},
  pages = {043052},
  numpages = {9},
  year = {2023},
  month = {10},
  publisher = {American Physical Society},
  doi = {10.1103/PhysRevResearch.5.043052},
  url = {https://link.aps.org/doi/10.1103/PhysRevResearch.5.043052}
}

@article{RoyMohseni2016,
url = {https://doi.org/10.1515/zkri-2016-2004},
title = {The Kepler tiling as the oldest complex surface structure in history: X-ray structure analysis of a two-dimensional oxide quasicrystal approximant},
author = {Sumalay Roy and Katayoon Mohseni and Stefan Förster and Martin Trautmann and Florian Schumann and Eva Zollner and Holger Meyerheim and Wolf Widdra},
pages = {749--755},
volume = {231},
number = {12},
journal = {Zeitschrift für Kristallographie - Crystalline Materials},
doi = {doi:10.1515/zkri-2016-2004},
year = {2016},
}

@book{kepler1997harmony,
  title        = {The Harmony of the World},
  author       = {Kepler, Johannes},
  editor       = {Aiton, E. J. and Duncan, A. M. and Field, J. V.},
  year         = {1997},
  publisher    = {American Philosophical Society},
  isbn         = {9780871692092},
  note         = {Translated into English with an introduction and notes by E.J. Aiton, A.M. Duncan, and J.V. Field},
  url          = {https://nla.gov.au/nla.cat-vn3994270},
}

@book{dresselhaus2007group,
  title={Group theory: application to the physics of condensed matter},
  author={Dresselhaus, Mildred S and Dresselhaus, Gene and Jorio, Ado},
  year={2007},
  publisher={Springer Science \& Business Media}
}

@article{hasan2010colloquium,
  title = {Colloquium: Topological insulators},
  author = {Hasan, M. Z. and Kane, C. L.},
  journal = {Rev. Mod. Phys.},
  volume = {82},
  issue = {4},
  pages = {3045--3067},
  numpages = {0},
  year = {2010},
  month = {11},
  publisher = {American Physical Society},
  doi = {10.1103/RevModPhys.82.3045},
  url = {https://link.aps.org/doi/10.1103/RevModPhys.82.3045}
}

@article{PhysRevLett.95.146802,
  title = {${Z}_{2}$ Topological Order and the Quantum Spin Hall Effect},
  author = {Kane, C. L. and Mele, E. J.},
  journal = {Phys. Rev. Lett.},
  volume = {95},
  issue = {14},
  pages = {146802},
  numpages = {4},
  year = {2005},
  month = {9},
  publisher = {American Physical Society},
  doi = {10.1103/PhysRevLett.95.146802},
  url = {https://link.aps.org/doi/10.1103/PhysRevLett.95.146802}
}

@article{PhysRevB.74.195312,
  title = {Time reversal polarization and a ${Z}_{2}$ adiabatic spin pump},
  author = {Fu, Liang and Kane, C. L.},
  journal = {Phys. Rev. B},
  volume = {74},
  issue = {19},
  pages = {195312},
  numpages = {13},
  year = {2006},
  month = {11},
  publisher = {American Physical Society},
  doi = {10.1103/PhysRevB.74.195312},
  url = {https://link.aps.org/doi/10.1103/PhysRevB.74.195312}
}

@article{soluyanov2011computing,
  title = {Computing topological invariants without inversion symmetry},
  author = {Soluyanov, Alexey A. and Vanderbilt, David},
  journal = {Phys. Rev. B},
  volume = {83},
  issue = {23},
  pages = {235401},
  numpages = {9},
  year = {2011},
  month = {6},
  publisher = {American Physical Society},
  doi = {10.1103/PhysRevB.83.235401},
  url = {https://link.aps.org/doi/10.1103/PhysRevB.83.235401}
}

@article{fu2007topological,
  title = {Topological insulators with inversion symmetry},
  author = {Fu, Liang and Kane, C. L.},
  journal = {Phys. Rev. B},
  volume = {76},
  issue = {4},
  pages = {045302},
  numpages = {17},
  year = {2007},
  month = {7},
  publisher = {American Physical Society},
  doi = {10.1103/PhysRevB.76.045302},
  url = {https://link.aps.org/doi/10.1103/PhysRevB.76.045302}
}

@article{PhysRevB.82.245412,
  title = {Tight-binding theory of the spin-orbit coupling in graphene},
  author = {Konschuh, Sergej and Gmitra, Martin and Fabian, Jaroslav},
  journal = {Phys. Rev. B},
  volume = {82},
  issue = {24},
  pages = {245412},
  numpages = {11},
  year = {2010},
  month = {12},
  publisher = {American Physical Society},
  doi = {10.1103/PhysRevB.82.245412},
  url = {https://link.aps.org/doi/10.1103/PhysRevB.82.245412}
}

@article{de2019topological,
  title={Topological flat band, Dirac fermions and quantum spin Hall phase in 2D Archimedean lattices},
  author={De Lima, F Crasto and Ferreira, Gerson J and Miwa, RH},
  journal={Physical Chemistry Chemical Physics},
  volume={21},
  number={40},
  pages={22344--22350},
  year={2019},
  publisher={Royal Society of Chemistry},
  doi={https://doi.org/10.1039/C9CP04760C}
}

@article{springer2020topological,
  title={Topological two-dimensional polymers},
  author={Springer, Maximilian A and Liu, Tsai-Jung and Kuc, Agnieszka and Heine, Thomas},
  journal={Chemical Society Reviews},
  volume={49},
  number={7},
  pages={2007--2019},
  year={2020},
  publisher={Royal Society of Chemistry},
  doi={https://doi.org/10.1039/C9CS00893D}
}

@article{PhysRevB.80.113102,
  title = {Topological insulator on the kagome lattice},
  author = {Guo, H.-M. and Franz, M.},
  journal = {Phys. Rev. B},
  volume = {80},
  issue = {11},
  pages = {113102},
  numpages = {4},
  year = {2009},
  month = {9},
  publisher = {American Physical Society},
  doi = {10.1103/PhysRevB.80.113102},
  url = {https://link.aps.org/doi/10.1103/PhysRevB.80.113102}
}

@article{RevModPhys.82.1539,
  title = {Anomalous Hall effect},
  author = {Nagaosa, Naoto and Sinova, Jairo and Onoda, Shigeki and MacDonald, A. H. and Ong, N. P.},
  journal = {Rev. Mod. Phys.},
  volume = {82},
  issue = {2},
  pages = {1539--1592},
  numpages = {0},
  year = {2010},
  month = {5},
  publisher = {American Physical Society},
  doi = {10.1103/RevModPhys.82.1539},
  url = {https://link.aps.org/doi/10.1103/RevModPhys.82.1539}
}

@article{RevModPhys.87.1213,
  title = {Spin Hall effects},
  author = {Sinova, Jairo and Valenzuela, Sergio O. and Wunderlich, J. and Back, C. H. and Jungwirth, T.},
  journal = {Rev. Mod. Phys.},
  volume = {87},
  issue = {4},
  pages = {1213--1260},
  numpages = {47},
  year = {2015},
  month = {10},
  publisher = {American Physical Society},
  doi = {10.1103/RevModPhys.87.1213},
  url = {https://link.aps.org/doi/10.1103/RevModPhys.87.1213}
}

@article{RevModPhys.82.1959,
  title = {Berry phase effects on electronic properties},
  author = {Xiao, Di and Chang, Ming-Che and Niu, Qian},
  journal = {Rev. Mod. Phys.},
  volume = {82},
  issue = {3},
  pages = {1959--2007},
  numpages = {0},
  year = {2010},
  month = {7},
  publisher = {American Physical Society},
  doi = {10.1103/RevModPhys.82.1959},
  url = {https://link.aps.org/doi/10.1103/RevModPhys.82.1959}
}

@article{PhysRevB.104.184423,
  title = {Spin Hall effect in noncollinear kagome antiferromagnets},
  author = {Busch, Oliver and G\"obel, B\"orge and Mertig, Ingrid},
  journal = {Phys. Rev. B},
  volume = {104},
  issue = {18},
  pages = {184423},
  numpages = {9},
  year = {2021},
  month = {11},
  publisher = {American Physical Society},
  doi = {10.1103/PhysRevB.104.184423},
  url = {https://link.aps.org/doi/10.1103/PhysRevB.104.184423}
}

@article{hongdeTriangulene,
	author = {Yu, Hongde and Heine, Thomas},
	doi = {10.1021/jacs.3c05178},
	journal = {Journal of the American Chemical Society},
	month = {09},
	number = {35},
	pages = {19303--19311},
	publisher = {American Chemical Society},
	title = {Magnetic Coupling Control in Triangulene Dimers},
	type = {doi: 10.1021/jacs.3c05178},
	url = {https://doi.org/10.1021/jacs.3c05178},
	volume = {145},
	year = {2023}
}

@article{hongdeTriangulene2,
	author = {Yu, Hongde and Wang, Dong},
	doi = {10.1021/jacs.0c02254},
	journal = {Journal of the American Chemical Society},
	month = {06},
	number = {25},
	pages = {11013--11021},
	publisher = {American Chemical Society},
	title = {Metal-Free Magnetism in Chemically Doped Covalent Organic Frameworks},
	type = {doi: 10.1021/jacs.0c02254},
	url = {https://doi.org/10.1021/jacs.0c02254},
	volume = {142},
	year = {2020},
}

@article{jonathanschmidtML,
	author = {Schmidt, Jonathan and Marques, M{\'a}rio R. G. and Botti, Silvana and Marques, Miguel A. L.},
	doi = {10.1038/s41524-019-0221-0},
	journal = {npj Computational Materials},
	number = {1},
	pages = {83},
	title = {Recent advances and applications of machine learning in solid-state materials science},
	url = {https://doi.org/10.1038/s41524-019-0221-0},
	volume = {5},
	year = {2019},
	}

@article{Gjerding_2021,
	author = {Morten Niklas Gjerding and Alireza Taghizadeh and Asbj{\o}rn Rasmussen and Sajid Ali and Fabian Bertoldo and Thorsten Deilmann and Nikolaj R{\o}rb{\ae}k Kn{\o}sgaard and Mads Kruse and Ask Hjorth Larsen and Simone Manti and Thomas Garm Pedersen and Urko Petralanda and Thorbj{\o}rn Skovhus and Mark Kamper Svendsen and Jens J{\o}rgen Mortensen and Thomas Olsen and Kristian Sommer Thygesen},
	doi = {10.1088/2053-1583/ac1059},
	journal = {2D Materials},
	month = {7},
	number = {4},
	pages = {044002},
	publisher = {IOP Publishing},
	title = {Recent progress of the Computational 2D Materials Database (C2DB)},
	url = {https://dx.doi.org/10.1088/2053-1583/ac1059},
	volume = {8},
	year = {2021}}

@article{Haastrup_2018,
doi = {10.1088/2053-1583/aacfc1},
url = {https://doi.org/10.1088/2053-1583/aacfc1},
year = {2018},
month = {sep},
publisher = {IOP Publishing},
volume = {5},
number = {4},
pages = {042002},
author = {Haastrup, Sten and Strange, Mikkel and Pandey, Mohnish and Deilmann, Thorsten and Schmidt, Per S and Hinsche, Nicki F and Gjerding, Morten N and Torelli, Daniele and Larsen, Peter M and Riis-Jensen, Anders C and Gath, Jakob and Jacobsen, Karsten W and Jørgen Mortensen, Jens and Olsen, Thomas and Thygesen, Kristian S},
title = {The Computational 2D Materials Database: high-throughput modeling and discovery of atomically thin crystals},
journal = {2D Materials}
}

@article{Haichenacs.jpclett.3c02131,
	author = {Wang, Hai-Chen and Huran, Ahmad W. and Marques, Miguel A. L. and Nalabothula, Muralidhar and Wirtz, Ludger and Romestan, Zachary and Romero, Aldo H.},
	doi = {10.1021/acs.jpclett.3c02131},
	journal = {The Journal of Physical Chemistry Letters},
	journal1 = {The Journal of Physical Chemistry Letters},
	journal2 = {J. Phys. Chem. Lett.},
	month = {11},
	number = {44},
	pages = {9969--9977},
	publisher = {American Chemical Society},
	title = {Two-Dimensional Noble Metal Chalcogenides in the Frustrated Snub-Square Lattice},
	type = {doi: 10.1021/acs.jpclett.3c02131},
	url = {https://doi.org/10.1021/acs.jpclett.3c02131},
	volume = {14},
	year = {2023},
	}

@article{wang2023symmetry,
  title={Symmetry-based computational search for novel binary and ternary 2D materials},
  author={Wang, Hai-Chen and Schmidt, Jonathan and Marques, Miguel AL and Wirtz, Ludger and Romero, Aldo H},
  journal={2D Materials},
  volume={10},
  number={3},
  pages={035007},
  year={2023},
  publisher={IOP Publishing},
  url={https://iopscience.iop.org/article/10.1088/2053-1583/accc43/meta},
}

@article{PhysRevLett.121.086602,
  title = {Intrinsic Spin and Orbital Hall Effects from Orbital Texture},
  author = {Go, Dongwook and Jo, Daegeun and Kim, Changyoung and Lee, Hyun-Woo},
  journal = {Phys. Rev. Lett.},
  volume = {121},
  issue = {8},
  pages = {086602},
  numpages = {6},
  year = {2018},
  month = {8},
  publisher = {American Physical Society},
  doi = {10.1103/PhysRevLett.121.086602},
  url = {https://link.aps.org/doi/10.1103/PhysRevLett.121.086602}
}

@article{PhysRevB.82.085310,
  title = {Topological insulators on the Lieb and perovskite lattices},
  author = {Weeks, C. and Franz, M.},
  journal = {Phys. Rev. B},
  volume = {82},
  issue = {8},
  pages = {085310},
  numpages = {5},
  year = {2010},
  month = {8},
  publisher = {American Physical Society},
  doi = {10.1103/PhysRevB.82.085310},
  url = {https://link.aps.org/doi/10.1103/PhysRevB.82.085310}
}

@article{FAN202330,
title = {Two-dimensional Dirac materials: Tight-binding lattice models and material candidates},
journal = {ChemPhysMater},
volume = {2},
number = {1},
pages = {30-42},
year = {2023},
issn = {2772-5715},
doi = {https://doi.org/10.1016/j.chphma.2022.04.009},
url = {https://www.sciencedirect.com/science/article/pii/S2772571522000298},
author = {Runyu Fan and Lei Sun and Xiaofei Shao and Yangyang Li and Mingwen Zhao}
}

@article{ase-paper,
  author={Ask Hjorth Larsen and Jens Jørgen Mortensen and Jakob Blomqvist and Ivano E Castelli and Rune Christensen and Marcin
Dułak and Jesper Friis and Michael N Groves and Bjørk Hammer and Cory Hargus and Eric D Hermes and Paul C Jennings and Peter
Bjerre Jensen and James Kermode and John R Kitchin and Esben Leonhard Kolsbjerg and Joseph Kubal and Kristen
Kaasbjerg and Steen Lysgaard and Jón Bergmann Maronsson and Tristan Maxson and Thomas Olsen and Lars Pastewka and Andrew
Peterson and Carsten Rostgaard and Jakob Schiøtz and Ole Schütt and Mikkel Strange and Kristian S Thygesen and Tejs
Vegge and Lasse Vilhelmsen and Michael Walter and Zhenhua Zeng and Karsten W Jacobsen},
  title={The atomic simulation environment—a Python library for working with atoms},
  journal={Journal of Physics: Condensed Matter},
  volume={29},
  number={27},
  pages={273002},
  url={http://stacks.iop.org/0953-8984/29/i=27/a=273002},
  year={2017}
}

@article{KaneMelePRL2005,
  author = {C. L. Kane and E. J. Mele},
  title = {Quantum Spin Hall Effect in Graphene},
  journal = {Phys. Rev. Lett.},
  volume = {95},
  pages = {226801},
  year = {2005},
  doi = {10.1103/PhysRevLett.95.226801}
}

@article{BernevigScience2006,
  author = {B. A. Bernevig and T. L. Hughes and S.-C. Zhang},
  title = {Quantum Spin Hall Effect and Topological Phase Transition in HgTe Quantum Wells},
  journal = {Science},
  volume = {314},
  pages = {1757--1761},
  year = {2006},
  doi = {10.1126/science.1133734}
}

@article{YeNature2018,
  author = {L. Ye and M. Kang and J. Liu and F. von Cube and C. R. Wicker and T. Suzuki and C. Jozwiak and A. Bostwick and E. Rotenberg and D. C. Bell and R. Comin and J. G. Checkelsky},
  title = {Massive Dirac fermions in a ferromagnetic kagome metal},
  journal = {Nature},
  volume = {555},
  pages = {638--642},
  year = {2018},
  doi = {10.1038/nature25987}
}

@article{YinNature2019,
  author = {J.-X. Yin and S. S. Zhang and G. Chang et al.},
  title = {Negative flat band magnetism in a spin–orbit-coupled correlated kagome magnet},
  journal = {Nature},
  volume = {562},
  pages = {91--95},
  year = {2019},
  doi = {10.1038/s41586-018-0502-7}
}

@book{SteinhardtOstlund1987,
  author = {P. J. Steinhardt and S. Ostlund},
  title = {The Physics of Quasicrystals},
  publisher = {World Scientific},
  year = {1987},
  doi = {https://doi.org/10.1142/0391}
}

@article{JAFFE1987399,
title = {Inclusion of spin-orbit coupling into tight binding bandstructure calculations for bulk and superlattice semiconductors},
journal = {Solid State Communications},
volume = {62},
number = {6},
pages = {399-402},
year = {1987},
issn = {0038-1098},
doi = {https://doi.org/10.1016/0038-1098(87)91042-8},
author = {M.D. Jaffe and J. Singh},
}

@misc{pythtb,
  author       = {Coh, Sinisa and
                  Vanderbilt, David},
  title        = {Python Tight Binding (PythTB)},
  month        = {9},
  year         = {2022},
  journal    = {Zenodo},
  doi          = {10.5281/zenodo.12721316},
  url          = {https://doi.org/10.5281/zenodo.12721316},
}

@article{gradhand2010,
  title = {Fully relativistic ab initio treatment of spin-flip scattering caused by impurities},
  author = {Gradhand, Martin and Fedorov, Dmitry V. and Zahn, Peter and Mertig, Ingrid},
  journal = {Phys. Rev. B},
  volume = {81},
  issue = {2},
  pages = {020403},
  numpages = {4},
  year = {2010},
  month = {1},
  publisher = {American Physical Society},
  doi = {10.1103/PhysRevB.81.020403},
  url = {https://link.aps.org/doi/10.1103/PhysRevB.81.020403}
}

@article{fabian99,
  title = {Phonon-Induced Spin Relaxation of Conduction Electrons in Aluminum},
  author = {Fabian, Jaroslav and Das Sarma, S.},
  journal = {Phys. Rev. Lett.},
  volume = {83},
  issue = {6},
  pages = {1211--1214},
  numpages = {0},
  year = {1999},
  month = {8},
  publisher = {American Physical Society},
  doi = {10.1103/PhysRevLett.83.1211},
  url = {https://link.aps.org/doi/10.1103/PhysRevLett.83.1211}
}

@Article{Kato2024,
author={Kato, Moyu
and Narumi, Yasuo
and Morita, Katsuhiro
and Matsushita, Yoshitaka
and Fukuoka, Shuhei
and Yamashita, Satoshi
and Nakazawa, Yasuhiro
and Oda, Migaku
and Hayashi, Hiroaki
and Yamaura, Kazunari
and Hagiwara, Masayuki
and Yoshida, Hiroyuki K.},
title={One-third magnetization plateau in Quantum Kagome antiferromagnet},
journal={Communications Physics},
year={2024},
month={12},
day={28},
volume={7},
number={1},
pages={424},
issn={2399-3650},
doi={10.1038/s42005-024-01922-0},
url={https://doi.org/10.1038/s42005-024-01922-0}
}

@article{PhysRevB.111.214412,
  title = {Magnetism of kagome metals $({\mathrm{Fe}}_{1\ensuremath{-}x}{\mathrm{Co}}_{x})\phantom{\rule{0.16em}{0ex}}\mathrm{Sn}$ studied by $\ensuremath{\mu}\mathrm{SR}$},
  author = {Cai, Yipeng and Yoon, Sungwon and Sheng, Qi and Zhao, Guoqiang and Seewald, Eric Francis and Ghosh, Sanat and Ingham, Julian and Pasupathy, Abhay Narayan and Queiroz, Raquel and Lei, Hechang and Xie, Yaofeng and Dai, Pengcheng and Ito, Takashi and Ke, Ruyi and Cava, Robert J. and Sharma, Sudarshan and Pula, Mathew and Luke, Graeme M. and Kojima, Kenji M. and Uemura, Yasutomo J.},
  journal = {Phys. Rev. B},
  volume = {111},
  issue = {21},
  pages = {214412},
  numpages = {17},
  year = {2025},
  month = {6},
  publisher = {American Physical Society},
  doi = {10.1103/PhysRevB.111.214412},
  url = {https://link.aps.org/doi/10.1103/PhysRevB.111.214412}
}

@article{PhysRevMaterials.5.034801,
  title = {Superconductivity in the ${\mathbb{Z}}_{2}$ kagome metal ${\mathrm{KV}}_{3}{\mathrm{Sb}}_{5}$},
  author = {Ortiz, Brenden R. and Sarte, Paul M. and Kenney, Eric M. and Graf, Michael J. and Teicher, Samuel M. L. and Seshadri, Ram and Wilson, Stephen D.},
  journal = {Phys. Rev. Mater.},
  volume = {5},
  issue = {3},
  pages = {034801},
  numpages = {7},
  year = {2021},
  month = {3},
  publisher = {American Physical Society},
  doi = {10.1103/PhysRevMaterials.5.034801},
  url = {https://link.aps.org/doi/10.1103/PhysRevMaterials.5.034801}
}

@article{Jovanovic_photonics,
author = {D. Jovanovi\'{c} and R. Gaji\'{c} and K. Hingerl},
journal = {Opt. Express},
number = {6},
pages = {4048--4058},
publisher = {Optica Publishing Group},
title = {Refraction and band isotropy in 2D square-like Archimedean photonic crystal lattices},
volume = {16},
month = {3},
year = {2008},
url = {https://opg.optica.org/oe/abstract.cfm?URI=oe-16-6-4048},
doi = {10.1364/OE.16.004048},
}

@article{Stelson_photonics,
    author = {Stelson, Angela C. and Britton, Wesley A. and Liddell Watson, Chekesha M.},
    title = {Photonic crystal properties of self-assembled Archimedean tilings},
    journal = {Journal of Applied Physics},
    volume = {121},
    number = {2},
    pages = {023101},
    year = {2017},
    month = {01},
    issn = {0021-8979},
    doi = {10.1063/1.4973472},
    url = {https://doi.org/10.1063/1.4973472}
}

@misc{SI, 
note = "See Supplemental Material at URL-will-be-inserted-by-publisher" }

@misc{zenodo_pham_hinsche_Mertig_2025,
  author       = {L. V. Duc Pham, Nicki F. Hinsche, Ingrid Mertig},
  title        = {Quantum spin hall phase in the truncated trihexagonal lattice: A topological Archimedean structure},
  month        = {12},
  year         = 2025,
  publisher    = {Zenodo},
  version      = {v3},
  doi          = {10.5281/zenodo.17956008},
  url          = {https://doi.org/10.5281/zenodo.17956008},
}

@article{Bechinger_quasiC,
	author = {Mikhael, Jules and Roth, Johannes and Helden, Laurent and Bechinger, Clemens},
	journal = {Nature},
	number = {7203},
	pages = {501--504},
	title = {Archimedean-like tiling on decagonal quasicrystalline surfaces},
	volume = {454},
	year = {2008}}

@article{GruznevTlBi,
	author = {Gruznev, Dimitry V. and Bondarenko, Leonid V. and Matetskiy, Andrey V. and Mihalyuk, Alexey N. and Tupchaya, Alexandra Y. and Utas, Oleg A. and Eremeev, Sergey V. and Hsing, Cheng-Rong and Chou, Jyh-Pin and Wei, Ching-Ming and Zotov, Andrey V. and Saranin, Alexander A.},
	date = {2016-01-19},
	doi = {10.1038/srep19446},
	isbn = {2045-2322},
	journal = {Scientific Reports},
	number = {1},
	pages = {19446},
	title = {Synthesis of two-dimensional TlxBi1-x compounds and Archimedean encoding of their atomic structure},
	url = {https://doi.org/10.1038/srep19446},
	volume = {6},
	year = {2016}
}

@article{Shi20242464,
title = {Quasicrystal approximants in isoreticular metal-organic frameworks via Cairo pentagonal tiling},
journal = {Chem},
volume = {10},
number = {8},
pages = {2464-2472},
year = {2024},
issn = {2451-9294},
doi = {https://doi.org/10.1016/j.chempr.2024.03.030},
author = {Le Shi and Zhangyi Xiong and Hao Wang and Honghao Cao and Zhijie Chen},
}

@article{VoigtQuasiC20,
	author = {Voigt, Laura and Kubus, Mariusz and Pedersen, Kasper S.},
	journal = {Nature Communications},
	number = {1},
	pages = {4705},
	title = {Chemical engineering of quasicrystal approximants in lanthanide-based coordination solids},
	volume = {11},
	year = {2020}}

@article{ChenMOF21,
author = {Chen, Hua and Voigt, Laura and Kubus, Mariusz and Mihrin, Dmytro and Mossin, Susanne and Larsen, Ren{\'e} W. and Kegnæs, Søren and Piligkos, Stergios and Pedersen, Kasper S.},
title = {Magnetic Archimedean Tessellations in Metal–Organic Frameworks},
journal = {Journal of the American Chemical Society},
volume = {143},
number = {35},
pages = {14041-14045},
year = {2021},
doi = {10.1021/jacs.1c05057}
}

@article{Niu2Dpoly,
author = {Niu, Tianchao and Hua, Chenqiang and Zhou, Miao},
title = {On-Surface Synthesis toward Two-Dimensional Polymers},
journal = {The Journal of Physical Chemistry Letters},
volume = {13},
number = {34},
pages = {8062-8077},
year = {2022},
doi = {10.1021/acs.jpclett.2c01481},
}

@article{YinXOFS,
	author = {Yin, Ruoting and Zhu, Xiang and Fu, Qiang and Hu, Tianyi and Wan, Lingyun and Wu, Yingying and Liang, Yifan and Wang, Zhengya and Qiu, Zhen-Lin and Tan, Yuan-Zhi and Ma, Chuanxu and Tan, Shijing and Hu, Wei and Li, Bin and Wang, Z. F. and Yang, Jinlong and Wang, Bing},
	date = {2024-04-06},
	doi = {10.1038/s41467-024-47367-5},
	id = {Yin2024},
	isbn = {2041-1723},
	journal = {Nature Communications},
	number = {1},
	pages = {2969},
	title = {Artificial kagome lattices of Shockley surface states patterned by halogen hydrogen-bonded organic frameworks},
	url = {https://doi.org/10.1038/s41467-024-47367-5},
	volume = {15},
	year = {2024},
}

@Article{Gomes2012,
author={Gomes, Kenjiro K. and Mar, Warren and Ko, Wonhee and Guinea, Francisco and Manoharan, Hari C.},
title={Designer Dirac fermions and topological phases in molecular graphene},
journal={Nature},
year={2012},
month={Mar},
day={01},
volume={483},
number={7389},
pages={306-310},
doi={10.1038/nature10941},
url={https://doi.org/10.1038/nature10941}
}

@article{GhaemiCOsurface,
  title = {Designer quantum spin Hall phase transition in molecular graphene},
  author = {Ghaemi, Pouyan and Gopalakrishnan, Sarang and Hughes, Taylor L.},
  journal = {Phys. Rev. B},
  volume = {86},
  issue = {20},
  pages = {201406},
  numpages = {5},
  year = {2012},
  month = {Nov},
  publisher = {American Physical Society},
  doi = {10.1103/PhysRevB.86.201406}
}

@misc{paupitz_review_allotr,
      title={A Concise Review of Recently Synthesized 2D Carbon Allotropes: Amorphous Carbon, Graphynes, Biphenylene and Fullerene Networks}, 
      author={Ricardo Paupitz and Alexandre F. Fonseca and Mizraim Bessa and Guilherme S. L. Fabris and William F. da Cunha and Leonardo D. Machado and Marcelo L. Pereira Junior and Luiz A. Ribeiro Junior and Douglas S. Galvão},
      year={2025},
      eprint={2509.01877},
      archivePrefix={arXiv},
      primaryClass={cond-mat.mtrl-sci},
      url={https://arxiv.org/abs/2509.01877}
}

@article{Zhang_kekulene,
	author = {Zhang, Zhenzhe and Pham, Hanh D. M. and Perepichka, Dmytro F. and Khaliullin, Rustam Z.},
	journal = {Nature Communications},
	number = {1},
	pages = {1953},
	title = {Prediction of highly stable 2D carbon allotropes based on azulenoid kekulene},
	volume = {15},
	year = {2024}}

@Article{Song_Graphenylene,
author ="Song, Qi and Wang, Bing and Deng, Ke and Feng, Xinliang and Wagner, Manfred and Gale, Julian D. and Müllen, Klaus and Zhi, Linjie",
title  ="Graphenylene{,} a unique two-dimensional carbon network with nondelocalized cyclohexatriene units",
journal  ="J. Mater. Chem. C",
year  ="2013",
volume  ="1",
issue  ="1",
pages  ="38-41",
publisher  ="The Royal Society of Chemistry",
doi  ="10.1039/C2TC00006G"
}

@article{DesyatkinGraphyne,
author = {Desyatkin, Victor G. and Martin, William B. and Aliev, Ali E. and Chapman, Nathaniel E. and Fonseca, Alexandre F. and Galvão, Douglas S. and Miller, Ericka Roy and Stone, Kevin H. and Wang, Zhong and Zakhidov, Dante and Limpoco, F. Ted and Almahdali, Sarah R. and Parker, Shane M. and Baughman, Ray H. and Rodionov, Valentin O.},
title = {Scalable Synthesis and Characterization of Multilayer gamma-Graphyne, New Carbon Crystals with a Small Direct Band Gap},
journal = {Journal of the American Chemical Society},
volume = {144},
number = {39},
pages = {17999-18008},
year = {2022},
doi = {10.1021/jacs.2c06583},
URL = {https://doi.org/10.1021/jacs.2c06583}
}

@article{FanBiphenylene,
author = {Qitang Fan  and Linghao Yan  and Matthias W. Tripp  and Ondřej Krejčí  and Stavrina Dimosthenous  and Stefan R. Kachel  and Mengyi Chen  and Adam S. Foster  and Ulrich Koert  and Peter Liljeroth  and J. Michael Gottfried },
title = {Biphenylene network: A nonbenzenoid carbon allotrope},
journal = {Science},
volume = {372},
number = {6544},
pages = {852-856},
year = {2021},
doi = {10.1126/science.abg4509},
URL = {https://www.science.org/doi/abs/10.1126/science.abg4509}
}

@Article{Lu_2D_allotropes,
author ="Lu, Haigang and Li, Si-Dian",
title  ="Two-dimensional carbon allotropes from graphene to graphyne",
journal  ="J. Mater. Chem. C",
year  ="2013",
volume  ="1",
issue  ="23",
pages  ="3677-3680",
publisher  ="The Royal Society of Chemistry",
doi  = {10.1039/C3TC30302K}
}

@Article{Li_graphdiyne,
author ="Li, Guoxing and Li, Yuliang and Liu, Huibiao and Guo, Yanbing and Li, Yongjun and Zhu, Daoben",
title  ="Architecture of graphdiyne nanoscale films",
journal  ="Chem. Commun.",
year  ="2010",
volume  ="46",
issue  ="19",
pages  ="3256-3258",
publisher  ="The Royal Society of Chemistry",
doi  = {10.1039/B922733D},
url  = {http://dx.doi.org/10.1039/B922733D}
}

@Article{Serafini_graphdiynes,
author ="Serafini, P. and Milani, A. and Tommasini, M. and Castiglioni, C. and Proserpio, D. M. and Bottani, C. E. and Casari, C. S.",
title  ="Vibrational properties of graphdiynes as 2D carbon materials beyond graphene",
journal  ="Phys. Chem. Chem. Phys.",
year  ="2022",
volume  ="24",
issue  ="17",
pages  ="10524-10536",
publisher  ="The Royal Society of Chemistry",
doi  ="10.1039/D2CP00980C"}

@ARTICLE{Chelnokov_photonics,
  author={David, S. and Chelnokov, A. and Lourtioz, J.-M.},
  journal={IEEE Journal of Quantum Electronics}, 
  title={Isotropic photonic structures: Archimedean-like tilings and quasi-crystals}, 
  year={2001},
  volume={37},
  number={11},
  pages={1427-1434},
  doi={10.1109/3.958365}}

@article{Valenzuela_2006,
	author = {Valenzuela, S. O. and Tinkham, M.},
	journal = {Nature},
	number = {7099},
	pages = {176--179},
	title = {Direct electronic measurement of the spin Hall effect},
	volume = {442},
	year = {2006}}

@article{KatoSHE,
author = {Y. K. Kato  and R. C. Myers  and A. C. Gossard  and D. D. Awschalom },
title = {Observation of the Spin Hall Effect in Semiconductors},
journal = {Science},
volume = {306},
number = {5703},
pages = {1910-1913},
year = {2004},
doi = {10.1126/science.1105514}
}

@article{Buhrman_spintorque,
author = {Luqiao Liu  and Chi-Feng Pai  and Y. Li  and H. W. Tseng  and D. C. Ralph  and R. A. Buhrman},
title = {Spin-Torque Switching with the Giant Spin Hall Effect of Tantalum},
journal = {Science},
volume = {336},
number = {6081},
pages = {555-558},
year = {2012},
doi = {10.1126/science.1218197},
URL = {https://www.science.org/doi/abs/10.1126/science.1218197},
}

@article{TangARPES,
	author = {Tang, Shujie and Zhang, Chaofan and Wong, Dillon and Pedramrazi, Zahra and Tsai, Hsin-Zon and Jia, Chunjing and Moritz, Brian and Claassen, Martin and Ryu, Hyejin and Kahn, Salman and Jiang, Juan and Yan, Hao and Hashimoto, Makoto and Lu, Donghui and Moore, Robert G. and Hwang, Chan-Cuk and Hwang, Choongyu and Hussain, Zahid and Chen, Yulin and Ugeda, Miguel M. and Liu, Zhi and Xie, Xiaoming and Devereaux, Thomas P. and Crommie, Michael F. and Mo, Sung-Kwan and Shen, Zhi-Xun},
	date = {2017-07-01},
	doi = {10.1038/nphys4174},
	isbn = {1745-2481},
	journal = {Nature Physics},
	number = {7},
	pages = {683--687},
	title = {Quantum spin Hall state in monolayer 1T'-WTe2},
	url = {https://doi.org/10.1038/nphys4174},
	volume = {13},
	year = {2017}
}

@article{UgedaSTM,
	author = {Ugeda, Miguel M. and Bradley, Aaron J. and Zhang, Yi and Onishi, Seita and Chen, Yi and Ruan, Wei and Ojeda-Aristizabal, Claudia and Ryu, Hyejin and Edmonds, Mark T. and Tsai, Hsin-Zon and Riss, Alexander and Mo, Sung-Kwan and Lee, Dunghai and Zettl, Alex and Hussain, Zahid and Shen, Zhi-Xun and Crommie, Michael F.},
	date = {2016-01-01},
	doi = {10.1038/nphys3527},
	isbn = {1745-2481},
	journal = {Nature Physics},
	number = {1},
	pages = {92--97},
	title = {Characterization of collective ground states in single-layer NbSe2},
	url = {https://doi.org/10.1038/nphys3527},
	volume = {12},
	year = {2016},
}

@article{WuARPESedge,
  title = {Evidence for Topological Edge States in a Large Energy Gap near the Step Edges on the Surface of ${\mathrm{ZrTe}}_{5}$},
  author = {Wu, R. and Ma, J.-Z. and Nie, S.-M. and Zhao, L.-X. and Huang, X. and Yin, J.-X. and Fu, B.-B. and Richard, P. and Chen, G.-F. and Fang, Z. and Dai, X. and Weng, H.-M. and Qian, T. and Ding, H. and Pan, S. H.},
  journal = {Phys. Rev. X},
  volume = {6},
  issue = {2},
  pages = {021017},
  numpages = {8},
  year = {2016},
  month = {May},
  publisher = {American Physical Society},
  doi = {10.1103/PhysRevX.6.021017},
  url = {https://link.aps.org/doi/10.1103/PhysRevX.6.021017}
}

@article{Joseph2025,
  title = {Walking on Archimedean lattices: Insights from Bloch band theory},
  author = {Joseph, Davidson Noby and Boettcher, Igor},
  journal = {Phys. Rev. E},
  volume = {112},
  issue = {4},
  pages = {044118},
  numpages = {29},
  year = {2025},
  month = {Oct},
  publisher = {American Physical Society},
  doi = {10.1103/1fvj-91v6},
  url = {https://link.aps.org/doi/10.1103/1fvj-91v6}
}

@article{PhysRevB.102.245427,
  title = {Band flattening in buckled monolayer graphene},
  author = {Milovanovi\ifmmode \acute{c}\else \'{c}\fi{}, S. P. and An\dj{}elkovi\ifmmode \acute{c}\else \'{c}\fi{}, M. and Covaci, L. and Peeters, F. M.},
  journal = {Phys. Rev. B},
  volume = {102},
  issue = {24},
  pages = {245427},
  numpages = {13},
  year = {2020},
  month = {Dec},
  publisher = {American Physical Society},
  doi = {10.1103/PhysRevB.102.245427},
  url = {https://link.aps.org/doi/10.1103/PhysRevB.102.245427}
}

@misc{vanpoppelen2025bucklingflatbandstwisted,
      title={Buckling and flat bands in twisted bilayer graphene}, 
      author={Jannes van Poppelen and Annica M. Black-Schaffer},
      year={2025},
      eprint={2510.13471},
      archivePrefix={arXiv},
      primaryClass={cond-mat.mes-hall},
      url={https://arxiv.org/abs/2510.13471}
}

@article{PhysRevB.111.235407,
  title = {Strain engineering of flat bands in buckled graphene superlattices},
  author = {Yuan, Xiaoyi and Zhang, Zichong and Zhu, Shuze},
  journal = {Phys. Rev. B},
  volume = {111},
  issue = {23},
  pages = {235407},
  numpages = {8},
  year = {2025},
  month = {Jun},
  publisher = {American Physical Society},
  doi = {10.1103/PhysRevB.111.235407},
  url = {https://link.aps.org/doi/10.1103/PhysRevB.111.235407}
}

@article{Lopez-Suarez_buckling,
author = {Lopez-Suarez, Miquel  and Lleopart, Genis  and Morales-Salvador, Raul  and Moreira, Ibério de P. R.  and Bromley, Stefan T. },
title = {Buckletronics: how compression-induced buckling affects the mechanical and electronic properties of sp<sup>2</sup>-based two-dimensional materials},
journal = {Philosophical Transactions of the Royal Society A: Mathematical, Physical and Engineering Sciences},
volume = {381},
number = {2250},
pages = {20220248},
year = {2023},
doi = {10.1098/rsta.2022.0248},
URL = {https://royalsocietypublishing.org/doi/abs/10.1098/rsta.2022.0248}
}

@article{AvsarSOC,
  author       = {Avsar, A. and Tan, J. Y. and Taychatanapat, T. and
                  Balakrishnan, J. and Koon, G. K. W. and Yeo, Y. and Lahiri, J. and
                  Carvalho, A. and Rodin, A. S. and O'Farrell, E. C. T. and
                  Eda, G. and Castro Neto, A. H. and {\''O}zyilmaz, B.},
  title        = {Spin-orbit proximity effect in graphene},
  journal      = {Nature Communications},
  date         = {2014-09-26},
  volume       = {5},
  pages        = {4875},
  doi          = {10.1038/ncomms5875},
  issn         = {2041-1723},
}

@article{PhysRevB.111.205415,
  title = {Increasing the proximity-induced spin-orbit coupling in bilayer $\mathrm{graphene}/{\mathrm{WSe}}_{2}$ heterostructures with pressure},
  author = {Szentp\'eteri, B\'alint and M\'arffy, Albin and Kedves, M\'at\'e and T\'ov\'ari, Endre and F\"ul\"op, B\'alint and K\"ukemezey, Istv\'an and Magyarkuti, Andr\'as and Watanabe, Kenji and Taniguchi, Takashi and Csonka, Szabolcs and Makk, P\'eter},
  journal = {Phys. Rev. B},
  volume = {111},
  issue = {20},
  pages = {205415},
  numpages = {7},
  year = {2025},
  month = {May},
  publisher = {American Physical Society},
  doi = {10.1103/PhysRevB.111.205415},
  url = {https://link.aps.org/doi/10.1103/PhysRevB.111.205415}
}

@article{SunSOC,
  author       = {Sun, Lihuan and Rademaker, Louk and Mauro, Diego and
                  Scarfato, Alessandro and P{\'a}sztor, {\'A}rp{\'a}d and
                  Guti{\'e}rrez-Lezama, Ignacio and Wang, Zhe and
                  Martinez-Castro, Jose and Morpurgo, Alberto F. and
                  Renner, Christoph},
  title        = {Determining spin--orbit coupling in graphene by quasiparticle interference imaging},
  journal      = {Nature Communications},
  date         = {2023-06-24},
  volume       = {14},
  pages        = {3771},
  doi          = {10.1038/s41467-023-39453-x},
  issn         = {2041-1723},
}

@article{FabianSOC,
  title = {Twist-angle dependent proximity induced spin-orbit coupling in graphene/transition metal dichalcogenide heterostructures},
  author = {Naimer, Thomas and Zollner, Klaus and Gmitra, Martin and Fabian, Jaroslav},
  journal = {Phys. Rev. B},
  volume = {104},
  issue = {19},
  pages = {195156},
  numpages = {13},
  year = {2021},
  month = {Nov},
  publisher = {American Physical Society},
  doi = {10.1103/PhysRevB.104.195156},
  url = {https://link.aps.org/doi/10.1103/PhysRevB.104.195156}
}

@article{WangGating,
	author = {Wang, Depeng and Zhao, Shufang and Yin, Ruiyang and Li, Linlin and Lou, Zheng and Shen, Guozhen},
	date = {2021-06-14},
	doi = {10.1038/s41528-021-00110-2},
	isbn = {2397-4621},
	journal = {npj Flexible Electronics},
	number = {1},
	pages = {13},
	title = {Recent advanced applications of ion-gel in ionic-gated transistor},
	url = {https://doi.org/10.1038/s41528-021-00110-2},
	volume = {5},
	year = {2021},
}

@article{KakenovGating,
author = {Kakenov, Nurbek and Balci, Osman and Takan, Taylan and Ozkan, Vedat Ali and Altan, Hakan and Kocabas, Coskun},
title = {Observation of Gate-Tunable Coherent Perfect Absorption of Terahertz Radiation in Graphene},
journal = {ACS Photonics},
volume = {3},
number = {9},
pages = {1531-1535},
year = {2016},
doi = {10.1021/acsphotonics.6b00240},
URL = {https://doi.org/10.1021/acsphotonics.6b00240}
}

@article{ChuGating,
	author = {Chu, Leiqiang and Schmidt, Hennrik and Pu, Jiang and Wang, Shunfeng and {\"O}zyilmaz, Barbaros and Takenobu, Taishi and Eda, Goki},
	date = {2014-12-03},
	doi = {10.1038/srep07293},
	id = {Chu2014},
	isbn = {2045-2322},
	journal = {Scientific Reports},
	number = {1},
	pages = {7293},
	title = {Charge transport in ion-gated mono-, bi- and trilayer MoS2 field effect transistors},
	url = {https://doi.org/10.1038/srep07293},
	volume = {4},
	year = {2014}
}

@article{Wang_2024,
doi = {WangZ2},
url = {https://doi.org/10.1088/2053-1583/ad3136},
year = {2024},
month = {mar},
publisher = {IOP Publishing},
volume = {11},
number = {2},
pages = {025033},
author = {Wang, Baokai and Zhou, Xiaoting and Hung, Yi-Chun and Lin, Yen-Chuan and Lin, Hsin and Bansil, Arun},
title = {High spin-Chern-number insulator in $\alpha$-antimonene with a hidden topological phase},
journal = {2D Materials}
}

\end{document}